# Do we always need a filter?

Tiancheng Li, Juan M. Corchado, Javier Bajo, Shudong Sun and Juan F. De Paz

BISITE research group, University of Salamanca, Spain; Northwestern Polytechnical University, China
tiancheng.li1985@gmail.com; corchado@usal.es; jbajo@fi.upm.es; sdsun@nwpu.eud.cn; fcofds@usal.es

## Abstract

Since the groundbreaking work of the Kalman filter in the 1960s, considerable effort has been devoted to various discrete time filters for dynamic state estimation, especially including dozens of different types of suboptimal implementations of the Bayes filters. This has been accompanied by the rapid development of simulation/approximation theories and technologies. The essence of filters is to make the best use of the observation information in time sequence based on the hidden Markov model, which however is advisable only under the premise that the modeling errors are relatively small and that the approximation used is not too much. While admitting the success of filters in many cases, this study investigates the failure cases when they are in fact ineffective for state estimation. Several classic models have shown that the straightforward observation-only ($O_2$) inference that does not need modeling the state dynamics can perform better (in terms of both accuracy and computing speed) for estimation than filters in certain cases. Special attention has been paid to quantitatively analyze when and why a filter will not outperform the $O_2$ inference from the information fusion perspective.

Thanks to the rapid development of advanced sensors and sensor network, the $O_2$ inference is not only engineering friendly and computationally fast but can also be very accurate and reliable by fusing the information received from multiple sensors. The statistical attributes of the multi-sensor $O_2$ inference are analyzed and demonstrated through simulations. In the situation with limited sensors, the $O_2$ approach can work jointly with existing clutter filtering and data association algorithms for multi-target tracking in clutter environments. Given an adequate number of sensors, the $O_2$ approach can employ the multi-sensor data fusion to deal with clutter and can handle the very general multi-target tracking scenario with no background information.

## Keywords:

Bayes filtering; target tracking; Kalman filter; particle filter; sensor fusion

Tiancheng Li, Juan M. Corchado, Javier Bajo and Juan, F. De Paz are with the BISITE research group, Faculty of Science, University of Salamanca, Salamanca 37008, Spain. (BISITE: http://bisite.usal.es)
Tiancheng Li (also) and Shudong Sun are with Northwestern Polytechnical University, Xi'an 710072, China. Javier Bajo is also with Department of Artificial Intelligence, Technical University of Madrid, Spain

T. Li will correspond to this work, email: tiancheng.li1985@gmail.com; t.c.li@{usal.es, mail.nwpu.edu.cn}
The living version of this document can be found at https://sites.google.com/site/tianchengli85/o2

## I.   Introduction: an example where filters fail

Dynamic state estimation, namely filtering, is concerned with the sequential process of estimating a state which evolves over time and which is periodically observed by sensors. This is often modeled as a hidden Markov model (HMM) where the system being modeled is assumed to be a Markov process with unobserved states, which can be either discrete-time or continuous-time. Usually, the discrete-time Markov process of the state transition is modelled as difference equation (1) and the continuous-time Markov process is modelled as differential equation (2).

$$x_t = f_t(x_{t-1}, u_t) \tag{1}$$

$$\frac{d}{dt}x_t = f_t(x_t, u_t) \tag{2}$$

where $t$ indicates the time instant (in Eq. (1), it is discrete positive integer while in Eq. (2) it is non-negative), $x_t$ denotes the state, and $u_t$ denotes the noise affecting the state transition/dynamics equation $f_t$. Often, $f_t$ can be time varying and is often unknown. In this work, we focus on the discrete time form given by Eq. (1) only.

In contrast, the observation is always received in discrete-time

$$y_t = h_t(x_t, v_t) \tag{3}$$

where $y_t$ denotes the observation (also called measurement), and $v_t$ denotes the noise affecting the observation equation $h_t$. $h_t$ reflects the sensor's working principle and is often known (and constant).

To estimate the state based on the noisy observations over time, a general method is to employ a filter based on the state space model (SSM) consists of Eq. (1) or (2) and (3). Bayesian filter forms the standard solution to the estimation problem, which consists of predicting (based on the state transition function) and correcting (using the observation information to correct the prediction) two steps such as, e.g., the Kalman filter and its later extensions, particle filters. The Kalman filter (KF), which is the closed form filtering solution to linear system with additive Gaussian noise (a special case of HMM), can be presented as one of the simplest dynamic Bayesian networks. KF and its extensions [1, 2, 3] calculate estimates of the true values of states (in terms of both Gaussian mean and variance) recursively over time, while the particle filter (PF) calculates estimates of the probability density function (PDF) of the state recursively over time, both using an assumed Markov process model (i.e. the state transition equation) and incoming observations in discrete time. In this document, both the variants of Kalman filters and particle filters will be investigated.

For simplicity, we start from a one-dimension SSM that has been widely employed for filter evaluation since first proposed in [4], with the state transition equation and the observation equation respectively given as follows

$$x_t = 1 + \sin(w\pi t) + \phi_1 x_{t-1} + u_t \tag{4}$$

$$y_t = \begin{cases} \phi_2 x_t^2 + v_t & t \leq 30 \\ \phi_3 x_t - 2 + v_t & t > 30 \end{cases} \tag{5}$$

where $x_t, y_t$ are the respective state and observation at time $t$, the scale parameters $\omega = 4\text{e} - 2$, $\phi_1 = 0.5$, $\phi_2 = 0.2$ and $\phi_3 = 0.5$, the process noise $u_t$ is a Gamma $\mathcal{G}a(3,2)$ random variable and the observation noise is Gaussian $v_t \sim \mathcal{N}(0, 0.00001)$. These are the default parameter setting in many publications including [4].

Intuitively, one of the simplest ways to estimate the state is to infer it directly from the observation. We call this method the observation-only (O₂) inference (that can also be referred to O₂I or O2I), which is filtering-free and non-Bayesian. Formally, and according to (3), the O₂ inference can be conceptually written as follows (as long as the observation function is invertible; see Section III.D):

$$\hat{x}_t = h_t^{-1}(y_t, v_t) \tag{6}$$

where $h_t^{-1}$ is the inverse function of $h_t$ in real variable space.

However, the inversing often introduces biases (the expectation of the estimate is not equal to the true state) if the function is nonlinear, where the bias highly depends on the noise and the nonlinearity; the larger the noise and the nonlinearity, the larger the bias/error. To a degree, the converting bias/error can be alleviated in an explicit form for simple know noises (such as Gaussian noises). Significantly different to filters, the O₂ inference here does not assume the observation noises and therefore it works for the case of unknown and even time-varying observation noises. Hence, we (have to) omit this issue when the noise $v_t$ is unknown by setting it to be zero. Then, Eq. (6) reduces to

$$\hat{x}_t = h_t^{-1}(y_t, \mathbf{0}) \tag{7}$$

For the observation function (5) with unknown noise, one has

$$\hat{x}_t = \begin{cases} +/- \sqrt{|y_t/\phi_2|} & t \leq 30 \\ \frac{y_t + 2}{\phi_3} & t > 30 \end{cases} \tag{8}$$

where the sign $+/-$ is problematic because the observation function is non-monotonic[1].

---

[1] The observation function here when $t \leq 30$ is a non-monotonic function and therefore the inversing calculation involves a sign problem as shown in the root calculation of Eq. (8) and will cause bias because of the noise (see Section III.B). If there is no prior information about the sign of the state (there are cases in which the state is bounded in a limited space with known sign), two ways are available to determine the sign of the estimate: the first is to estimate it based on the state transition function and the previous estimate, namely the sign function: $\text{sgn}(\hat{x}_t) = \text{sgn}(f_t(\hat{x}_{t-1}))$. The second way is to use an additional sign estimator (see our simulation given in Section IV.A), which is computationally more expensive. In both methods, the O₂ method will not only explore the observation information but also somewhat the state transition knowledge. Here we use the first way to determine the sign of the estimate as default. There are possible other efficient methods to deal with this but here, we emphasize the value of the estimate.

Clearly, the $O_2$ inference directly explores the observations for estimating regardless of the noise contained (as no filtering is implemented) and is a deterministic method. If the noise is known, debiasing technologies shall be applied, for which we propose a Monte Carlo simulation method in Section III.B. Furthermore, the observation function may be irreversible or may involves under/over-determination and incomplete-estimation issues which will be detailed in Section III.D. Furthermore, a simply analysis of the Fisher efficiency of the $O_2$ inference will be given in section III.E.

Since the observation noise is zero-mean with a small variance in this case, the $O_2$ result is unbiased and might be comparable to a filter. To verify this, a series of typical filters are employed for comparison with the $O_2$ inference that includes the Extended Kalman filter (EKF) [2, 3], the Unscented Kalman filter (UKF) [13], the bootstrap filter (PF) [5] and the particle filters that uses EKF and UKF separately as the proposal, namely EKPF and UKPF (see [4] for detail). For the particle filters, we use 200 particles. For the EKF/UKF, we use the initial state variance of 0.75. The unscented transform parameter is set as $\alpha = 1, \beta = 0, \kappa = 2$ (the same as given in [4]). The true state and the initial unbiased estimate of all filters all start from $x_1 = 1$. Since EKF/UKF cannot be used directly for this Gamma noise, we assume equivalent variance 0.75 as an alternative for use (admitting a modeling error/bias of 1.5, as $\mathcal{G}a(3,2)$ is of mean 1.5, variance 0.75).

To compare the estimate accuracy of different filters, the root mean square error (RMSE) is used and defined on the time series as follows

$$\text{RMSE}_t = \sqrt{\frac{1}{M}\sum_{i=1}^{M}(x_{t,i} - \hat{x}_{t,i})^2} \tag{9}$$

where $M$ is the number of MC runs, $x_{t,i}$ is the true state at time $t$ of run $i$ and $\hat{x}_{t,i}$ is the state-estimate. To capture the average result, 200 MC runs are performed with random re-initialization for each run. Each run consists of 60 time-steps.

The true state and estimates given by different filters for one run are plotted in Fig.1. The RMSE of different methods are plotted in Fig.2. The mean and variance of RMSE over time and the computing time of each method are given in Table I. It is a "surprise" that the $O_2$ method has outperformed all the filters by several orders of magnitude in terms of both RMSE and computing speed, which indicates that the prediction-correction filters (at least those that have been used) are ineffective and unnecessary for this model. Further results on this model will be given at the end of Section IV.A for different observation noises (see Fig.18). According to our knowledge, such a good result has never been reported on any filter although many have been proposed to apply on this model. Then, why does the simplest $O_2$ method perform the best? Will the same result occur to other models? What is the core different between the filter and the $O_2$ inference?

*A filter shall only be applied when it can at minimum improve the estimation of the $O_2$ inference, although it can hardly be faster than the $O_2$ inference*. The above simulation simply indicates that the filters might be in fact ineffective in certain cases (the observation noise is very small in this case), although among the different filter options one might be better than another. It is critical to distinguish these cases so that one is clear whether to use a filter or just $O_2$ inference for a particular case.

In the next section, we will study the problem under the most typical Gaussian models. It is interesting and also critical to find that fusing two sources of information, which is the core art of the filter, may not obtain a better estimate than using only one of the two.

Table I Performance of different filters and the $O_2$ inference

|  | RMSE | | Computing time |
|---|---|---|---|
|  | mean | variance |  |
| EKF | 0.343 | 0.161 | 0.004 |
| UKF | 0.271 | 0.096 | 0.017 |
| SIR(PF) | 0.572 | 0.071 | 0.947 |
| EKPF | 0.345 | 0.161 | 1.988 |
| UKPF | 0.231 | 0.070 | 4.941 |
| $O_2$ Inference | 0.006 | $1.05 \times 10^{-5}$ | $1.09 \times 10^{-5}$ |

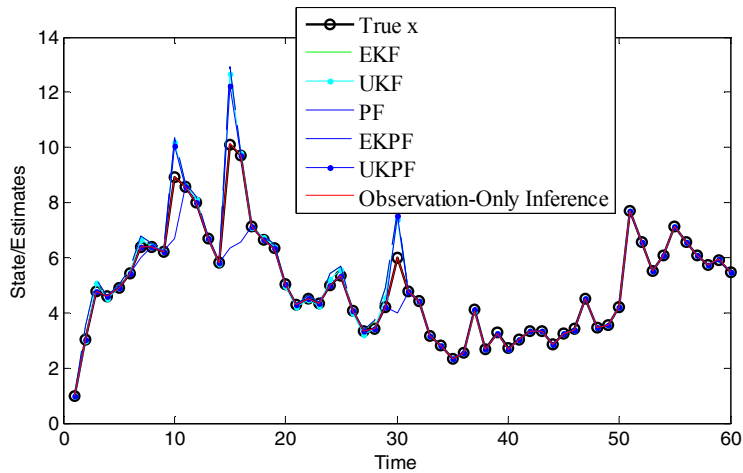

Fig.1 The true state and estimates of different filters and the $O_2$I method

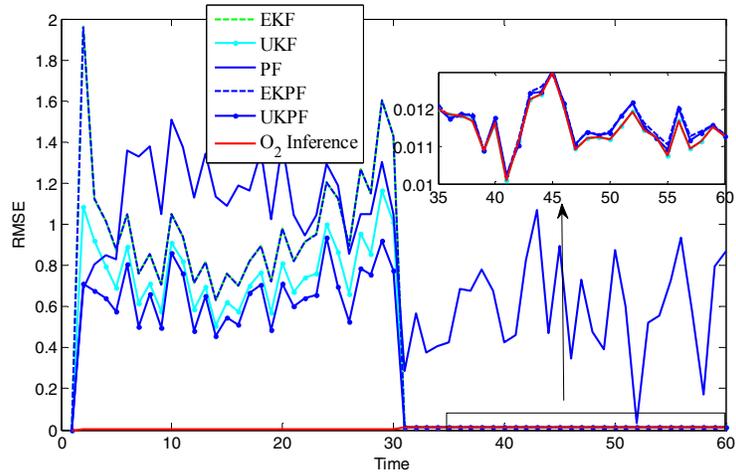

Fig.2 RMSE of different filters and the $O_2$I method

The remainder of this document is organized as follows. Section II investigates certain cases to discuss the probability that the filter may (/not) benefit the state estimation and the reason. Based on this, Section III summaries the findings and discussions on the effectiveness of the filter and the $O_2$ inference. Section IV presents more quantitative simulation evidence based on representative problem models. Finally, Section V presents the conclusions obtained based on our findings.

## II An information fusion view of the Bayes filter

Basically speaking, discrete-time recursive filters (typically including Kalman filters and particle filters) comprise two steps: prediction and correction (also called updating). The prediction is using the state transition equation to infer a priori estimate $p(x_{t|t-1})$ (that is a Markov process) based on the previous posterior while the updating step uses the observation to correct the prediction (according to KF rule, PF rule, etc.), obtaining a posterior estimate of the state $p(x_{t|t})$. This involves information fusion of two distributions: the prediction distribution and the observation-inference (also be referred to as the so-called "likelihood" distribution). Particularly, the well-known Kalman filter [1, 2, 3, 13, 14] gives the optimal fusion of two Gaussian distributions (corresponding the prediction and the observation) that minimizes the square estimate error. For more complicated distributions, the case will be more complicated (e.g. the particle Bayes filter which is based on random sampling) but they share the same story.

In the following we assume they are Gaussian, either biased or unbiased with regard to the true state. An analysis is provided from the information fusion perspective to check whether the posterior estimate given by the fusion of them (by using KF rule or PF rule) will be better (closer to the true state) than the estimate inferred from the observation only. If yes, the filter is beneficial otherwise the filter is just useless. Our discussion does not intend to seamlessly cover all the cases but to expose a core part of the problem.

Given two Gaussian distributions $p(x) = \mathcal{N}(m_x, \delta_x^2)$, $p(y) = \mathcal{N}(m_y, \delta_y^2)$ and the true state $m_T$, the goal of the filter is fusing $p(x)$ and $p(y)$ to get a combined distribution $p(z) = \mathcal{N}(m_z, \delta_z^2)$ as an estimate of the true state $m_T$. Using the KF rule that fuses data according to covariance, one has

$$m_z = \frac{\delta_x^2 m_y + \delta_y^2 m_x}{\delta_x^2 + \delta_y^2} \tag{10}$$

$$\delta_z^2 = \frac{\delta_x^2 \delta_y^2}{\delta_x^2 + \delta_y^2} \tag{11}$$

However, we will show in the following subsections that $z \sim p(z)$ might not be a more accurate estimate than $x \sim p(x)$ or $y \sim p(y)$ i.e. $|m_T - z|$ is not guaranteed to smaller than $|m_T - x|$ or $|m_T - y|$, although the variance of the estimate will be surely reduced as $\delta_z^2 \leq \{\delta_x^2, \delta_y^2\}$ which means a reduction of the uncertainty of the estimate.

Here, mapping from the observation space to the state space is required to obtain a state distribution from the observation. Regardless the possible nonlinear inversing bias, this may magnify/minify the noise by mapping the observation noise to the state space. Therefore, we must be aware that even if the observation noise is small/large, its mapping in the state space might be large/small. Specifically, for the simplicity of denotation, the symbols $x, y, z$ used in this section refer to the estimate of the state (using different sources of information) and the variables are *scalar* (in the 1D state space).

### II.A  Case 1: KF-fusion of two unbiased Gaussian distributions

**Remark 1**: *Kalman filter-fusion of two unbiased Gaussian estimates gives an unbiased estimate with a variance smaller than any original Gaussian estimate*.

As shown in Fig.3, the KF fusion of the Gaussian distribution (blue) $p(x) = \mathcal{N}(0,400)$ and the distribution (green) $p(y) = \mathcal{N}(0,100)$ gets a distribution (red) $p(z) = \mathcal{N}(0,80)$ according to Eq. (10~11), where the variance $\delta_z^2 = 80$ is smaller both than $\delta_x^2 = 400$ and $\delta_y^2 = 100$. Assuming $p(x)$ is inferred from observation, the KF-fusion will reduce the estimate variance by $\delta_x^2 - \delta_z^2$. It can be calculated that the estimate $z \sim p(z)$ is approximately 73.2% possibility by Eq. (12) better than $x \sim p(x)$ in the sense that $|m_T - x| > |m_T - z|$. In this case, the filter/fusion will give a more accurate estimate with a smaller variance, i.e. the estimate $z \sim p(z)$ is more possibly to be closer to the true state than $x \sim p(x)$ or $y \sim p(y)$ (see study given in the following subsection and Fig.5 for $p = |m_y - m_x| = 0$).

For a range of different variance ratio $\delta_y^2/\delta_x^2$, the probability of $P|m_T - x| > |m_T - z|$ is given by the red curve in Fig.5. This is the case in which the filter will be helpful (as the PoFB is always larger than 50%).

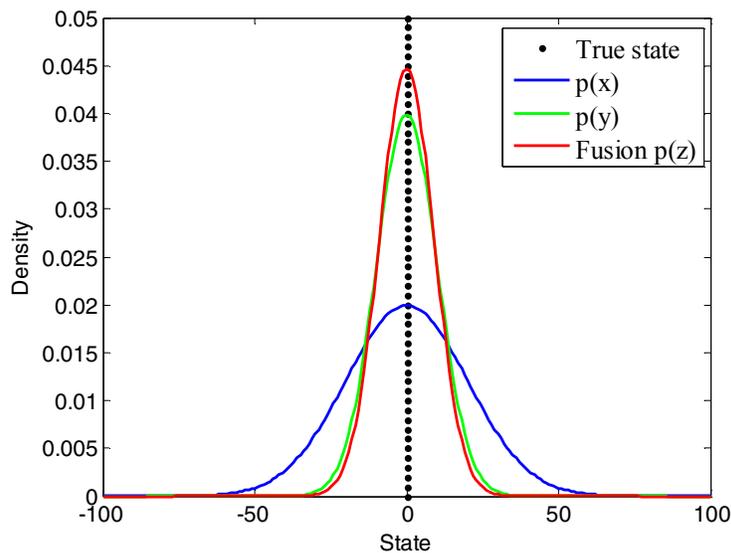

Fig.3   Fusion of two unbiased Gaussian distributions

T. Li *et al.* Do we always need a filter? arXiv: 1408.4636However, it is rare to have both distributions unbiased in the prediction-correction filters; instead, one or both of them can be biased (explanation is given in Section III.A), which is studied in the following two cases.

## II.B  Case 2: KF-fusion of one biased and one unbiased Gaussian distribution

**Remark 2**: *Kalman filter-fusion of one unbiased Gaussian estimate with one biased Gaussian estimate gives an almost surely biased estimate with a mean between the means of the two original Gaussian estimates, and with a variance smaller than both variances of the two original Gaussian estimates.*

As shown in Fig.4, the KF fusion of the Gaussian distribution (blue) $p(x) = \mathcal{N}(0,400)$ and the distribution (green) $p(y) = \mathcal{N}(50,100)$ gives a distribution (red) $p(z) = \mathcal{N}(40,80)$ according to (10~11). We assume that the distribution of $p(x)$ is unbiased (i.e. $m_T = m_x$) and the distribution (green) $p(y) = \mathcal{N}(50,100)$ is biased with the bias $|m_y - m_x|$. In this case, the estimate given by the fusion distribution $z \sim p(z)$ is not guaranteed to be better than the estimate $x \sim p(x)$ given by the unbiased distribution, although it will surely be better than the biased estimate $y \sim p(y)$. We can study the case in more detail. It is necessary to note that if there is only one distribution among the prediction and observation in a filter that is unbiased, it must be the observation. This is because the observation is independent of the prediction but the prediction dependent on the observation. A biased observation will surely cause a biased prediction for the next time-instant but the contrary does not hold. Therefore, in this case, the unbiased estimate $x \sim p(x)$ is inferred from the observation and the biased $p(y)$ is the prediction.

Denoting $x$ the estimate given by $p(x)$ and $z$ the estimate given by $p(z)$, we have $x \sim p(x), z \sim p(z)$. It is when and only when $|m_T - x| > |m_T - z|$, that estimate $z$ is better than $x$. The probability of fusion benefit (PoFB) can be calculated as follows

$$P(|m_T - x| > |m_T - z|)$$
$$= P((m_T - x)^2 > (m_T - z)^2)$$
$$= P((2m_T - z - x)(z - x) > 0)$$
$$= P\big((2m_T - z - x) > 0, (z - x) > 0\big) + P\big((2m_T - z - x) < 0, (z - x) < 0\big)$$
$$= P\big(x < z < (2m_T - x)\big) + P(2m_T - x < z < x) \tag{12}$$

It is known that the cumulative distribution function of the Gaussian distribution $p(z) = \mathcal{N}(m_z, \delta_z^2)$ is

$$\Phi_z(z) = \frac{1}{\delta_z \sqrt{2\pi}} \int_{-\infty}^{z} e^{-(t-m_z)^2/2\delta_z^2} \, dt \tag{13}$$

Therefore, Eq. (12) is equivalent to

$$\int_{-\infty}^{m_T} \big(\Phi_z(2m_T - x) - \Phi_z(x)\big) p(x) dx + \int_{m_T}^{\infty} \big(\Phi_z(x) - \Phi_z(2m_T - x)\big) p(x) dx \tag{14}$$

where $p(x) = \frac{1}{\delta_x \sqrt{2\pi}} e^{-(x-m_x)^2/2\delta_x^2}$.

Similarly, we have

$$P(|m_T - y| > |m_T - z|) = P(y < z < (2m_T - y)) + P(2m_T - y < z < y)$$
$$= \int_{-\infty}^{m_T} (\Phi_z(2m_T - y) - \Phi_z(y))p(y)dy + \int_{m_T}^{\infty} (\Phi_z(y) - \Phi_z(2m_T - y))p(y)dy \quad (15)$$

where $p(y) = \frac{1}{\delta_y \sqrt{2\pi}} e^{-(y-m_y)^2/2\delta_y^2}$.

Further, we define variance ratio (VR) $r$, the ratio of the variances of two distributions, and bias ratio (BR) $p$, the ratio of the distance between the means of two distributions over the standard deviation of the unbiased distribution respectively as

$$r = \frac{\delta_y^2}{\delta_x^2} \quad (16)$$

$$p = \frac{m_y - m_x}{\delta_x} \quad (17)$$

The result of Eq. (14/15) is highly related to VR $r$ and BR $p$. Due to the symmetry of the Gaussian distribution, we only consider the case of a positive BR $p \geq 0$ and the result holds the same for a negative BR.

To have a clear understanding of the PoFB, 100,000 samples are generated separately from distributions $x \sim p(x)$, $y \sim p(y)$ and $z \sim p(z)$ to calculate PoFB by comparing the fusion estimate $z$ to $x, y$ respectively for different VR $r \in [0.01, 1000]$ and BR $p \in [0,10]$. In particular, $p = 0$ means both distributions are unbiased (i.e. Case 1 given in section II.A). The results of Eq. (14) and (15) are shown in Fig.5 and Fig. 6 respectively. Fig. 6 shows that the fusion is always more likely to get a better estimate than the biased prediction. Specifically, the larger $p$ is, the larger the PoFB is. We are more interested in the comparison between the observation-based inference $x \sim p(x)$ and the fusion $z \sim p(z)$, which is shown in Fig.5, and have made the following observations.

First, PoFB = $P(|m_T - x| > |m_T - z|; m_x = m_T)$ will tend to be stable with 0.5 when $r$ goes to infinite. In particular, for $p \geq 2$, the larger VR $r$, approximately the larger the PoFB is; for $p \leq 0.4$, the larger VR is, approximately the smaller the PoFB is; for $0.4 < p < 2$, the PoFB goes up (and pass 0.5) and then reduce down to 0.5 with the increasing of VR $r$. This agrees with the KF rule that a larger $r$ corresponds to a larger $\delta_y^2$ of $p(y)$ which will have a lesser effect on the fusion distribution $p(z)$. Therefore, for a very large $r$, the fusion effect can be omitted, leaving us with $p(z) = p(x)$, then estimate $z \sim p(z)$ vs $x \sim p(x)$ is 50-50.

Secondly, when the bias $p \leq 0.6$ (for $r \geq 0.01$), the fusion estimate $z \sim p(z)$ has more than an approximately 50% possibility of obtaining a more accurate estimate than the unbiased $x \sim p(x)$. This means that when the bias of the prediction is not significant, the fusion will be preferable and is more likely to benefit. *This is the case (when the prediction is only slightly biased) in which filters are still recommended.*

Thirdly, when the bias $p \geq 1$, the fusion estimate $z \sim p(z)$ has less than an approximately 50% possibility of obtaining a more accurate estimate than the unbiased $x \sim p(x)$. This means that when the bias of the prediction is significant, the fusion will be more likely to obtain a worse result than the unbiased observation-only estimate. *This is the case (the bias of the prediction is significant) in which filters are not recommended.*

As shown in the left subfigure of Fig.5, for $r \to 0$, the prediction $p(y)$ is extremely accurate (with extremely small variance; but biased) and will dominate the fusion result, leaving us with $p(z) = p(y)$, then PoFB = $P(|m_T - x| > |m_T - z|; m_x = m_T)$ will almost fully depend on the bias of the prediction $p$: the smaller $p$ is, the larger the PoFB is. However, in general the prediction of a filter that fuses both the process noise and the observation noise cannot be so accurate (as compared with the observation).

The results also indicate that if the prediction is slightly biased or unbiased ($p$ is relatively small), a small variance will be very helpful; otherwise, a small variance can be a disaster for fusion (e.g. when $r < 0.1, p > 2$, the PoFB is very low and close to zero). This means that a very accurate estimate of small variance, biased or unbiased, is a double-edged sword for an unbiased estimate in the Kalman-filer fusion. (For this, we have **Remark 4** given in Section III) This can explain why a filtering system of very accurate state dynamics model and very small process noise is not reliable/robust (very sensitive for system disturbances).

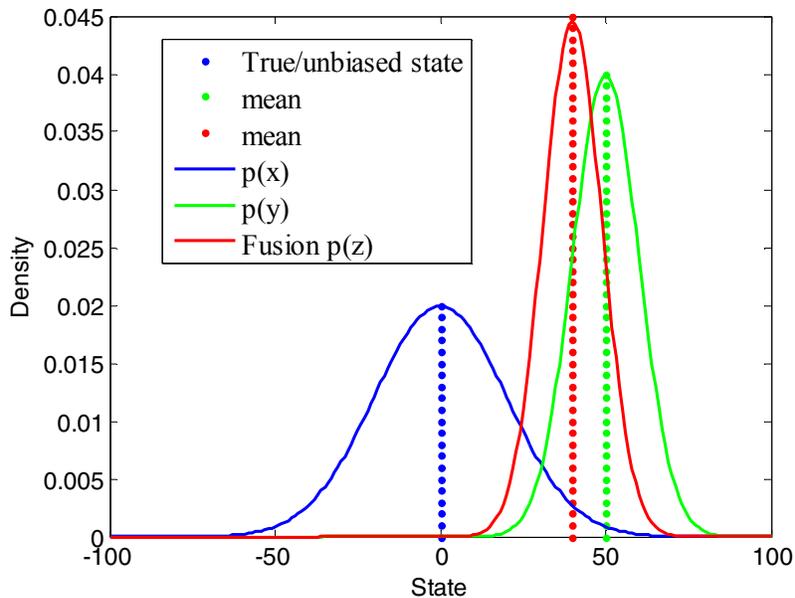

Fig. 4   Fusion of one unbiased distribution with one biased distribution

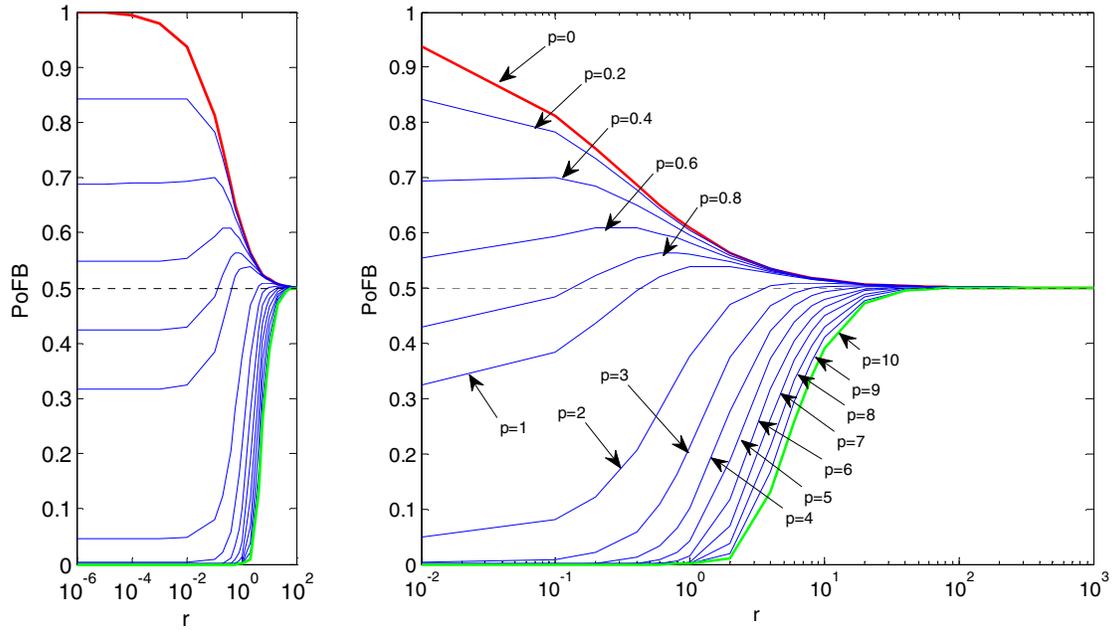

Fig.5    PoFB = $P(|m_T - x| > |m_T - z|; m_x = m_T)$ for different VR $r$ and bias $p$

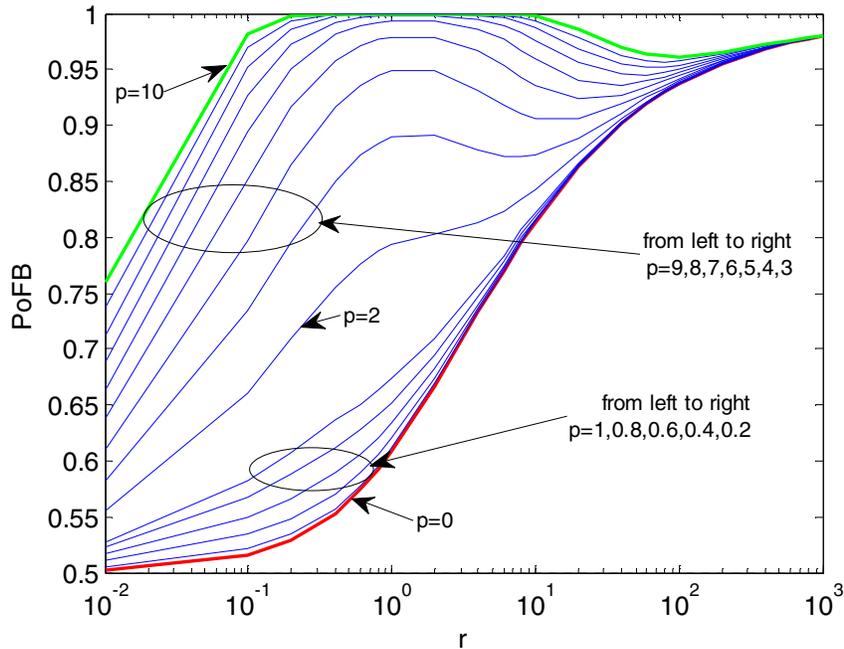

Fig.6    PoFB = $P(|m_T - y| > |m_T - z|; m_x = m_T)$ for different VR $r$ and bias $p$

## II.C    Case 3: KF-fusion of two biased Gaussian distributions

This is the case in which both the observation and the prediction are biased. For both, the bias is possible unknown. We present the following remark:

**Remark 3**: *Kalman filter-fusion of two biased Gaussian estimates gives an almost surely biased estimate with a smaller variance than any variance of the original Gaussian estimates. The bias of the fused estimate can be bigger or smaller than any bias of the original Gaussian estimates*.

This can be illustrated in Fig.7 which however only gives one specific case in which the blue distribution $p(x)$ is left biased (i.e. the mean of the estimate is on the left of the true state) and the green distribution $p(y)$ is right biased (i.e. the mean of the estimate is on the right of the true state), while the true state (black) is between them. This is the case (the bias of two distributions is in a specified direction and of a specified relative magnitude) in which the fusion $p(z)$ more possibly obtains a better estimate. In fact, the true state and the bias of the distributions are generally unknown; often left or right is 50-50. We will discuss several different cases.

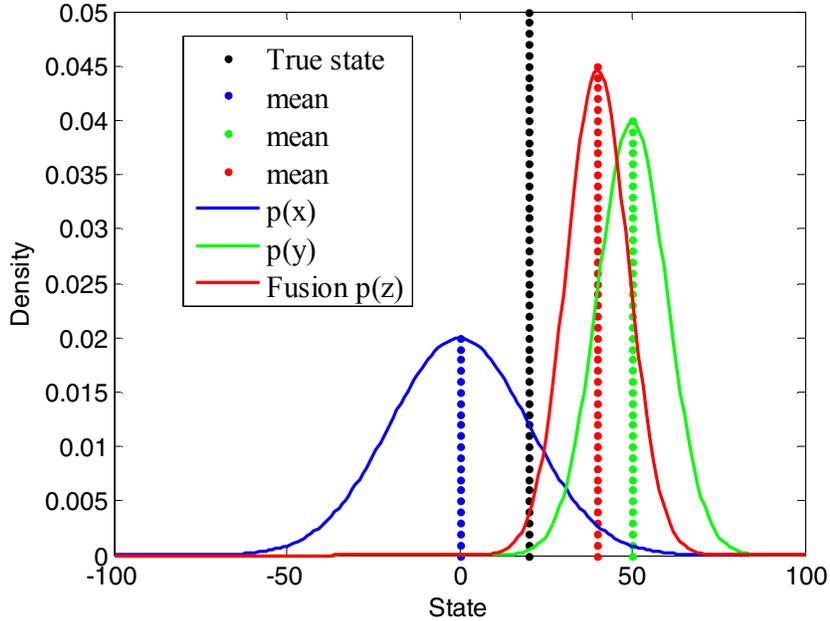

Fig. 7　Fusion of two biased Gaussian distributions

Denoting the estimate given by $p(x)$ as $x$, the estimate given by $p(y)$ as $y$ and the estimate given by $p(z)$ as $z$, we have $x \sim p(x), y \sim p(y), z \sim p(z)$. It is when and only when $\min(|m_T - x|, |m_T - y|) > |m_T - z|$, that estimate $z$ is better than $x$ and $y$. Obviously, the probability, namely the PoFB, is

$$P(\min(|m_T - x|, |m_T - y|) > |m_T - z|)$$
$$= P((m_T - x)^2 > (m_T - z)^2, (m_T - y)^2 > (m_T - z)^2)$$
$$= P(x < z < (2m_T - x), z < (2m_T - y)) + P(2m_T - x < z < x, 2m_T - y < z) \quad (18)$$
$$= \int_{-\infty}^{m_T} \int_{-\infty}^{x} (\Phi_z(2m_T - x) - \Phi_z(x)) p(x) p(y) dx dy$$

$$+ \int_{m_T}^{\infty} \int_{x}^{\infty} \left(\Phi_z(x) - \Phi_z(2m_T - x)\right) p(x) p(y) dx dy \tag{19}$$

Eq. (18/19) treats $x \sim p(x)$ and $y \sim p(y)$ equally. In the following we consider the case $m_y > m_x$ for simplicity; the results hold the same for $m_x > m_y$ due to the symmetry of the equation. 100,000 samples are drawn separately from the distribution $x \sim p(x)$, $y \sim p(y)$ and $z \sim p(z)$ to calculate the probability of Eq. (18) for a different true state $m_T$ which is chosen by adjusting a scaling parameter $m$

$$m = \frac{m_T - m_x}{\delta_x} \tag{20}$$

In the simulation, we use $m = \{-10, -5, -2, -1, -0.1, 0.1, 1, 2, 5, 10, 30\}$ separately. The results are shown in Fig.8 for different values of $m, r$ and $p$ ($r$ and $p$ are defined as Eq. (16) and (17)). Each subfigure corresponds to a different value of $m$ and each line in each subfigure corresponds to a different $p$ (the red line is for $p = 0$, the green line is for $p = 10$ while the blue lines are between).

First, all PoFBs seem to converge to 50% when the $r$ goes to infinite. In more detail, the PoFB is almost surely smaller than 50% when $m_T \leq m_x$ and when $m_T \geq m_y$. Only in the case when $m_x < m_T < m_y$ (or $m_x < m_T < m_y$ according to the symmetry of Eq. (18)), which corresponds to the case that the true state $m_T$ is between the two means of $p(x)$ and $p(y)$, is there more than a 50% possibility that the fusion can benefit the estimate, i.e. PoFB> 0.5. More precisely, a larger PoFB is more likely to be obtained when $m_T$ happens to be close to $m_z$ (that needs a proper configuration between $m, r$ and $p$). It is necessary to note that when the true state $m_T$ is between the means of $p(x)$ and $p(y)$, the performance order of the filters with different $p$ has a change. This is because Eq. (18) is a non-monotonic function due to the minimizing calculation.

There are more perspectives to understand the results. We are mainly interested in the results showing it is less likely to get a better estimate by fusing two biased estimates than the better of either one, except that the true state happens to lie around a proper position between two biased estimates (which cannot be guaranteed at all). However, it is generally unknown which of the distribution $p(x)$ and $p(y)$ is better in practice. In addition, if we don't apply a filter, we don't have a prediction and only the observation-based inference $p(x)$ is feasible to use. Therefore, it is more preferable to compare the observation-based inference $x \sim p(x)$ (if filter does not apply) with the fusion $z \sim p(z)$ (a filter applies) as done in Eq. (12) to see whether it is worthwhile/rewarding to employ a filter. For different $m$, the results of (12) are plotted separately in Fig.9.

Again, all PoFBs will converge to 50% when the $r$ goes into infinite. The results given in Fig.9 further show that: when $m \leq 0$ (i.e. $m_T \leq m_x \leq m_y$), all PoFBs will be smaller than 50% and the larger $p$ is, the smaller the PoFB is, while when $m \geq p$ (i.e. $m_x \leq m_y \leq m_T$), all PoFBs will be larger than 50% and the larger $p$ is, the larger the PoFB is.

When $0 < m < p$ (i.e. $m_x < m_T < m_y$), the PoFB depends on $r, m, p$ (see the sub-plots for $m = 1, 2, 5$): generally, with the increase of $r > 1$, the PoFB will go up and then go down to 0.5 finally. The primary indication is that when the true state $m_T$ and the mean of the prediction $m_y$ are at the same side of the mean of the likelihood distribution $m_x$ and also $p < m$, then the fusion is guaranteed to benefit. This is not practical for the general dynamic state estimation models. In practice, it is impossible to control the prediction to be at the same side as the true state (as the true state is never known). Right or left is 50-50. Overall, the filter is not so optimistic as one may expect to outperform the O$_2$ inference in the case that both observation and prediction are biased.

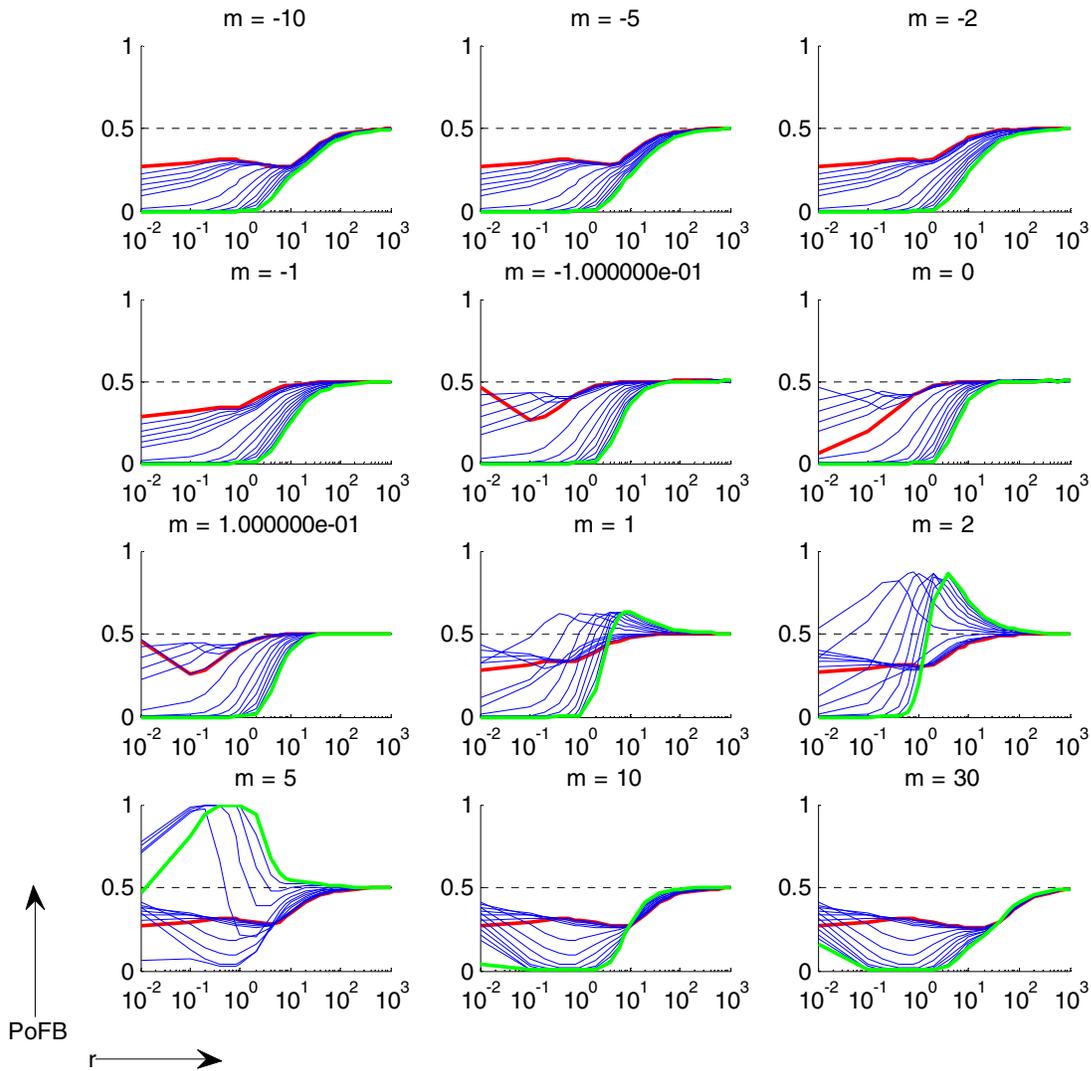

Fig. 8    PoFB $= P(\min(|m_T - x|, |m_T - y|) > |m_T - z|)$ for different true state $m_T$, VR $r$ and bias $p$; the red line is for $p = 0$, the green line is for $p = 10$ while the blue is between.

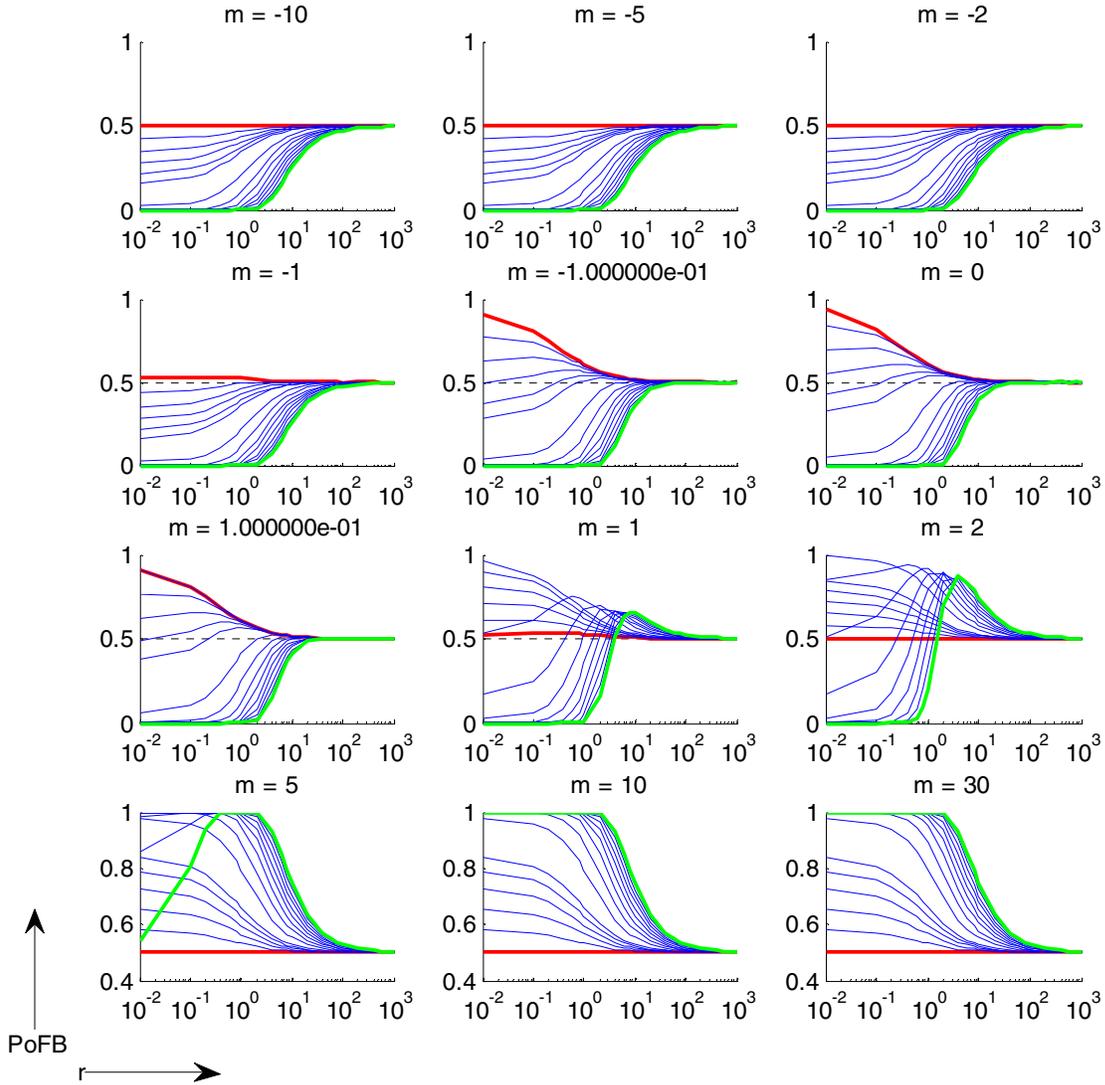

Fig. 9    PoFB = P($|m_T - x| > |m_T - z|$) for different true state $m_T = m \times \delta_x$, variance ration $r$ and bias $p$; the red line is for $p = 0$, the green line is for $p = 10$ while the blue lines are between

## II.D    Case 4: Particle Bayes filter-fusion

In this category, we will apply a particle method (typically such as the particle filter [3-7], the point mass method [8] and particle flow [9]) to represent the Gaussian distribution and to implement the prediction-correction fusion in a standard Bayes rule. The Bayes estimation of the state distribution can be expressed in terms of the filtering distribution at time instant $t-1$, $p(x_{t-1}|y_{1:t-1})$, the prediction distribution $p(x_t|x_{t-1})$ and the observation likelihood distribution $p(y_t|x_t)$ that is, in a recursive form by

$$p(x_t|y_{1:t}) \propto \int p(y_t|x_t)p(x_t|x_{t-1})p(x_{t-1}|y_{1:t-1})dx_{t-1} \tag{21}$$

where the symbol $\propto$ signifies "proportional to." This update cannot be implemented analytically except in a very few cases, and therefore one resorts to approximations.

The core idea of the PF [3-7] is to represent continuous distributions by a set of weighted particles $\{x_t^{(n)}, w_t^{(n)}\}, n = 1,2,\dots,N$ where particle $x_t^{(n)}$ are possible values of the unknown state $x_t$, $w_t^{(n)}$ are weights assigned to the particles, $N$ is the number of particles. Namely,

$$p(x_t) \approx \sum_{n=1}^{N} w_t^{(n)} \delta\left(x_t - x_t^{(n)}\right) \tag{22}$$

where $\delta(\cdot)$ is the Dirac delta impulse and all the weights sum up to one. When a new observation comes, the weights of the particles have to be reweighted based on the sequential importance sampling that lies in the Bayes updating rule

$$w_t^{(n)} \propto w_{t-1}^{(n)} \frac{p\left(y_t|x_t^{(n)}\right) p\left(x_t^{(n)}|x_{t-1}^{(n)}\right)}{q\left(x_t^{(n)}\right)} \tag{23}$$

where $q(\cdot)$ is a proposal distribution to generate particles. The bootstrap filter [5], also known as the basic particle filter, utilizes $q\left(x_t^{(n)}\right) = p\left(x_t^{(n)}|x_{t-1}^{(n)}\right)$ while EKPF/UKPF etc. utilize the EKF/UKF etc. to construct advanced proposal distribution.

Often, the computation of the expression to the right of the proportionality sign is followed by normalization of the weights (so that they sum up to one)

$$w_k^{(n)} = \frac{w_k^{(n)}}{\sum_n^N w_k^{(n)}} \tag{24}$$

After weight updating, resampling is often required to reduce the weight variance so that all particles will have an exactly equal or approximate weight. The resampling should not (significantly) change the distribution of particles, and shall usually be unbiased [10].

The particle method does not require the underlying distribution to be Gaussian. However, for simplicity, we still use the representative Gaussian distributions. Since the particle method itself is a Monte Carlo method, there are few clues to calculate the analytical probability that the fusion is better than the single estimate (especially when the resampling step is employed), as we did in the KF fusion of the previous three cases; however, we can easily use the sampling method to numerically simulate the probability of fusion benefit as defined in Eq. (12).

We directly assume $p(x)$ the distribution of the state inferred by the observation while $p(y)$ is the prediction represented by using particles $\{x_t^{(n)}, \frac{1}{N}\}, n = 1,2\dots N$ which are equally weighted, namely $p(y) \approx \sum_{n=1}^{N} \frac{1}{N} \delta\left(x_t - x_t^{(n)}\right)$. The correction uses the information contained in $p(x)$ to update the weights of all of the particles $\{w_t^{(n)}\}, n =$ 1,2 ... $N$ and then get the posterior distribution $p(z) \approx \sum_{n=1}^{N} w_t^{(n)} \delta\left(x_t - x_t^{(n)}\right)$. To save space, we will focus on the particle fusion result as compared to the $O_2$ method only, i.e. PoFB = $P(|m_T - x| > |m_T - z|)$, where $m_T$ is the true state.

First, we assume the observation-inference distribution $p(x)$ is unbiased. (We iterate that if there is only one distribution between the prediction and observation that is unbiased, it must be the observation). 1,000,000 samples are generated separately from the observation-inference distribution $x \sim p(x)$ and the fusion distribution $z \sim p(z)$ to calculate the probability $P(|m_T - x| > |m_T - z|)$ for different $r \in [0.01, 1000]$ and $p \in [0,10]$, where we use the same definition Eq. (16), (17) and (20) for $r, p$, and $m$. In particular, $p = 0$ means the prediction/prediction is also unbiased. The results are shown in Fig.10, which is very similar to the KF fusion as shown in Fig.5. This makes sense as, theoretically, if the variables are linear and normally distributed the Bayes filter becomes equal to the Kalman filter. The slight difference is due to the approximation of the particles, which is different from the closed form Kalman filter. We have the same primary observations as follows:

1) PoFB will tend to be stable with 0.5 when $r$ goes to infinite. Approximately, we have for $p \geq 2$, the larger $r$ is, the larger the PoFB is; for $p \leq 0.4$, the larger $r$ is, the smaller the PoFB is.

2) When the bias $p \leq 0.5$, the fusion estimate $z \sim p(z)$ has approximately more than 50% possibility of obtaining a more accurate estimate than the unbiased $x \sim p(x)$. Thus, when the bias of the prediction is not significant, filters are recommendable.

3) When the bias $p \geq 1$, the fusion estimate $p(z)$ has less than 50% possibility of obtaining a more accurate estimate than the unbiased $p(x)$. This means that when the bias of the predictionis significant, filters are not recommended.

Secondly, we set the observation distribution biased, i.e. $m_T \neq m_x$. For different $m$, the results are plotted separately in Fig.11. The results are again similar to the Kalman filter-type fusion as shown in Fig.9. the PoFB will converge to 50% when $r$ goes to infinite.

1) When $m \leq 0$ (i.e. $m_T \leq m_x \leq m_y$), all PoFBs will be smaller than 50% and the larger $p$ is, the smaller PoFB is;
2) When $m \geq p$ (i.e. $m_x \leq m_y \leq m_T$), all PoFBs will be larger than 50% and the larger $p$ is, the larger PoFB is;
3) When $0 < m < p$ (i.e. $m_x < m_T < m_y$), the PoFB depends on $r, m, p$: generally, with the increase of $r > 1$, the PoFB will go up (within a scope, the PoFB passes 0.5 significantly) and then go down to 0.5 finally.

Again, the filter is not optimistic to outperform the $O_2$ inference in the case that both observation and prediction are biased.

T. Li *et al*. Do we always need a filter? arXiv: 1408.4636

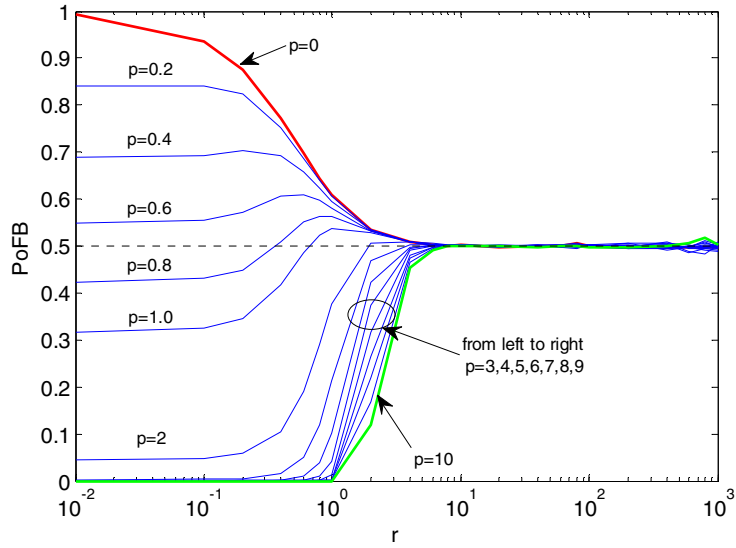

Fig.10　PoFB = $P(|m_T - x| > |m_T - z|; m_x = m_T)$ for different variance ration $r$ and bias $p$

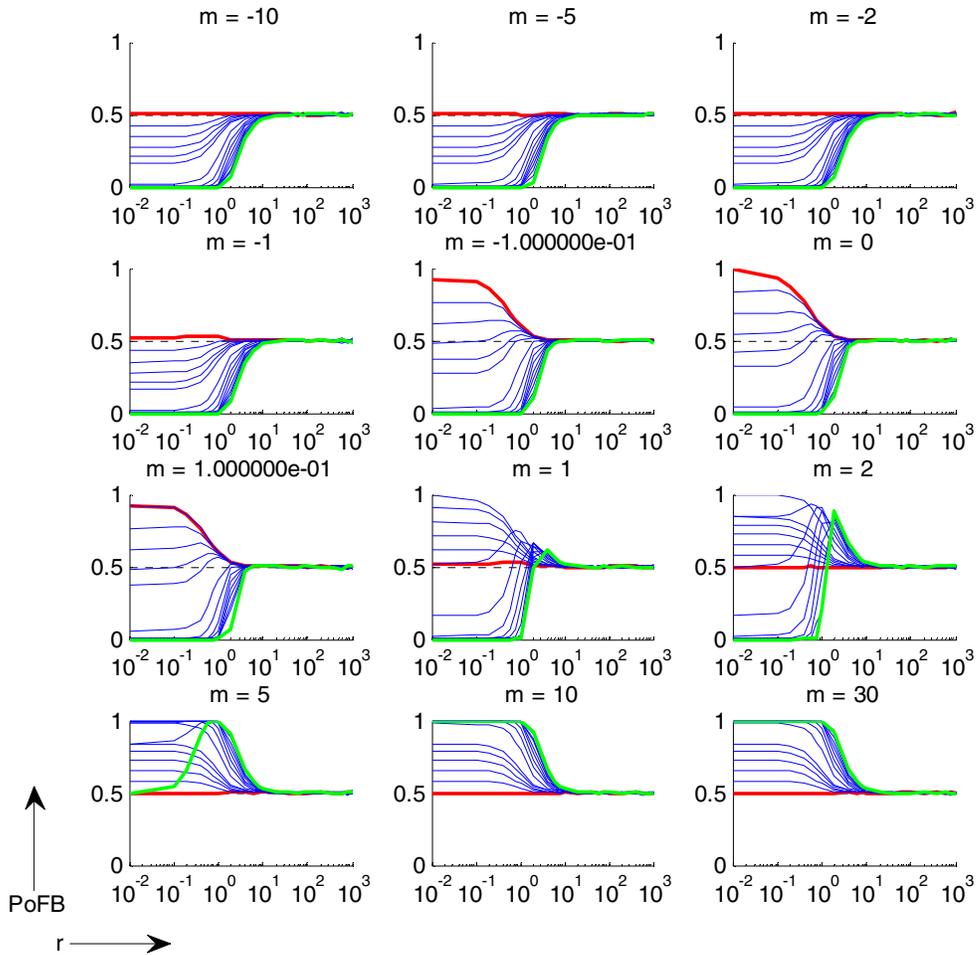

Fig. 11　PoFB = $P(|m_T - x| > |m_T - z|)$ for different true state $m_T = m \times \delta_x$, VR $r$ and bias $p$; the red line is for $p = 0$, the green line is for $p = 10$ while the blue lines are between

## III  Discussion: O$_2$ or filters?

The above statistical investigation shows that it is only in 1) the full **Case 1** (both observation and prediction are unbiased), 2) a part of **Case 2** (the bias of the prediction is very small while the observation is unbiased) and 3) a part of **Case 3** (the observation is much *worse* than the prediction) that the prediction-observation fusion, namely the filter, is likely to get a better estimate than the O$_2$ inference. Otherwise, the O$_2$ inference is more likely to perform better. Here we may have a general conclusion.

**Remark 4** *Whether the filter will outperform the unbiased O$_2$ inference or not primarily depends on the quality of the prediction, especially the bias of the prediction (the lesser the bias, the better); how much the benefit will be depends on the variance of the prediction (as compared to the variance of the observation).*

Generally, many issues affect the quality of the prediction as it has absorbed all the historical information including possible initialization error, system disturbance, bad observation, modeling error (including assumption on the process noise [11]). These together with the approximation used in the suboptimal filters will cause discrepancies (namely error/bias) between the prediction and the true states; this is the main reason why the correction is required, as the name suggests. This will be further discussed in the next subsection. Overall, the prediction is generally biased.

In contrast, the O$_2$ inference does not involve these modeling/approximation issues since it only requires the observation function $h_t$ that depends on the sensor's working principle and is often known in real life. As a general assumption on the sensors, the observation is often treated as unbiased. Even if there is sensor bias (e.g. register error), it shall be corrected offline in a way. But the user is hard to test the bias online since the observation is the only information that the user can trust. The observation cannot correct itself, which is the same in the filters. Therefore, this document will not discuss the rare case that the observation function is unknown, which is the same challenging for all estimators and has to be estimated before state estimation.

The O$_2$ inference in fact lies in the core of many wireless positioning technologies such as time difference of arrival (TDOA) techniques (the most typical example is GPS, global positioning system), signal strength methods and angle of arrival location, just to name a few. More straightforwardly, it was demonstrated that simple deterministic algorithms outperform the particle filter in a type of finite-state estimation [37], even given that the filter is provided with correct system modeling.

However, we must be aware that the fusion discussed so far maximally corresponds to one prediction-correction iteration of a discrete-time filter, while in the time-sequence, the condition of the system varies. That is to say, $r, p$ and $m$ vary with time, which can generate a situation at some stages filtering is better at some stages (the prediction obtained is good enough) while at some other stages (the prediction is relatively poor) it is not as good as the $O_2$ inference. It is desirable albeit challenging to switch them in real-time so that an "optimal" decision is made to allow $O_2I$ and filters to work interactively. Nevertheless, we have several general principles before we discuss further.

1) If the observation noise is significant or even not zero-mean, neither the $O_2I$ nor the filter can be good (comparably the $O_2I$ is more sensitive to the observation noise).

2) If the system can be correctly simulated/modeled, the filter can be well initialized and is affected with a relatively small process noise, the filter will work well.

3) If the state model cannot be correctly simulated (or the filter has to make great approximation) and there are many disturbances (or miss-detection) from time to time, the filter will not work well but instead it might be better to use $O_2$ rather than a filter.

4) At the initialization stage of a filter, the observation information can be explored to avoid large initialization error for a filter.

5) *A filter shall only be applied, whether in simulations or in real-life problems, when it at least outperforms the $O_2$ inference.*

### III.A  Use of prediction and observation

It is known that the Kalman filter under the linear system with additive Gaussian noises reaches the Cramér-Rao bound and is optimal. To illustrate this, we rewrite the simulation models given in Section I of Eq. (3~4) as follows

$$\begin{cases} x_t = 1 + \sin(0.04\pi t) + 0.5x_{t-1} + u_t \\ \quad\quad y_t = 0.5x_t - 2 + v_t \end{cases} \quad (25)$$

where noises are zero-mean Gaussian $u_t \sim \mathcal{N}(0,0.75), v_t \sim \mathcal{N}(0,R)$. For this linear and Gaussian SSM, the Kalman filter is directly applicable. For a range of variances $R$ from 0.00001 to 100, the average RMSE of the Kalman filter and the $O_2$ inference over 1000 Monte Carlo runs (every run consists of 1000 steps) are given in Fig.12.

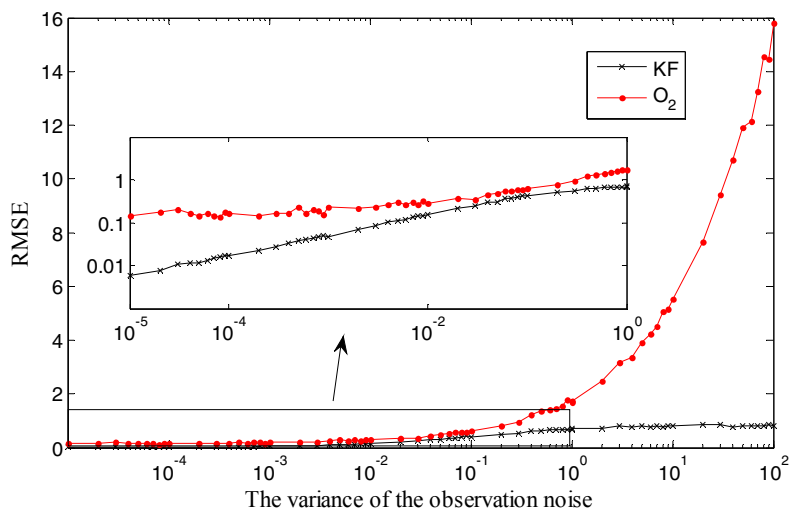

Fig. 12 Kalman filter outperforms the $O_2$ inference under exactly known linear and Gaussian system

Fig.12 shows that the Kalman filter does perform better than the $O_2$ inference whether the observation noise is large or small, because the filter just correctly assumes the real system, i.e. the filter knows exactly the system models and the level of Gaussian noises. For a large observation noise, the advantage of filtering is obvious as it can use the prediction to bind the observation while the RMSE of $O_2$ inference will grow unboundedly with the variance of the observation noise. In this linear observation model, it is linear growth. This is the advantage of the filter over the $O_2$ approach and a case where the filter is highly useful and recommended.

*There is much difference between such a correctly assumed simulations (the models of the true state/noise and the one used by the filter are exactly the same, which is suitable to apply directly an optimal filter) and real-life problems.*

In many situations the model of a real process may differ from those of the best available model of that process, we refer to this difference as modeling error, especially in the realistic case that the only information that is available (e.g. pedestrian tracking and weather forecast) is taken from the observations. What is used in the filter as the state transition model is only an assumption/simulation (referred to as *modeling*) which is not guaranteed to be exactly correct. The core of the filter is using a state transition model (given it is correctly assumed) to "propagate" the history information to fuse with the current observation, in order to make the best utilization of the data and the knowledge of the system. However, the accuracy of the models/knowledge used makes differences.

1) In the real world both the state transition and process noise vary and are often impossible to model accurately. One typical example involves tracking pedestrians where it is almost impossible to get an accurate model for the movement of human, regardless of process noises, unexpected system disturbance, missing observation.

2) It is often impossible to initialize a filter without introducing any bias except when the system is fully known in advance.

3) Nonlinearity prevents the direct application of the optimal filter that has to be approximated [3-9, 13-15, 20] at the price of approximation errors. This will reduce the quality of the prediction, rendering a reduction of the estimation accuracy.

4) Any error/bias, whether due to mismodeling or approximation, introduced into the posterior will be propagated to the following steps and will not be fully removed.

5) *With the joint application of multiple sensors, the observation obtained can be very accurate (corresponding to a large VR r in the discussion of Section II), preventing the necessity of the use of a filter, especially in realistic complicated systems where modeling error occurs always, to some content.*

All of these indicate that the discrete time filter is not really reliable but in fact significantly sensitive to modeling error, system disturbance and approximation error.

**Remark 5**. *Any modeling-based estimator/filter suffers from modeling error; the more the assumption/approximation is, the more unreliable the estimator/filter is.*

The sensitivity of the *filter* to the *model* is well reflected in the difference between the probability hypothesis density (PHD) filter and the multi-target multi-Bernoulli filter. They have obviously different approximation equations of the Bayes filter for the same multi-target tracking problem simply because of using different models of new-target appear process [17] only. What a worse result will be obtained if the real new-target appear model matches none of them?

In fact, the sensitivity of the filter to modeling error has already been acknowledged and corresponding treatments have been investigated as early as in the late 1960s, see e.g. [33, 29]. Recently, state space augmentation [34] and model assessment [30, 35] have been investigated. The strength of the aforementioned EKPF/UKPF that use EKF/UKF as the proposal is driving the prediction to match the unbiased observation (this is only efficient when the observation noise is relatively small; see the simulation results given in Fig. 17 and 18 in Section IV.A). One can use other algorithms to do the same thing (as long as the observation is unbiased and of small variance), which has become a common idea to improve the particle filter, see e.g. [3-7, 12] (we wonder whether this is still a rigorous Bayes filter). Similarly, the core idea of some "adaptive" (see e.g. [19]), "robust" (see e.g. [21, 33, 34, 36]) and "sparse" (see e.g. [36]) filtering and optimization techniques is to emphasize the observation information to improve the posterior distribution, see also [3, 20]. We will not detail these contexts here. So far, continuous efforts are still being devoted to design more advanced Bayes filters. Simply, bad information is detrimental for fusion and therefore should be avoided (at least not used so much). However, it is still unclear how to control or even to know the quality of the prediction in filters online.

It is clear that as long as the prediction is of good quality (unbiased or slightly biased), the prediction-observation fusion, namely the filter, will be effective for state estimation otherwise it is not guaranteed at all. *Nevertheless, even the state transition information is not so accurate to be useful in the filter fusion manner, the estimator may still benefit from it in another way.* In fact, we have already shown that a filter can be helpful to estimate the sign of the state for the $O_2$ inference. Moreover, existing data association or clutter-filtering algorithms (such as joint probability density association, multiple hypotheses tracker or the probability hypotheses density filter, etc. [16, 17]) can be applied with the $O_2$ inference. We will show how to combine the filter with the $O_2$ inference within the multi-object tracking content in our simulation of Section IV.B.

In short, as long as there is any useful information available about the state transition model (even if one cannot benefit from it in the manner of using a prediction-correction filter), it shall be useful for the $O_2$ inference (can be termed as $O_2+$), e.g. using it to determine the sign of the estimate, to infer the unobserved dimension of the state, to filter clutter, to distinguish estimates from one another if there are multiple objects [22] and to predict/smooth the estimate, etc. Therefore, we have the following remark.

T. Li *et al*. Do we always need a filter? arXiv: 1408.4636**Remark 6** *The $O_2+$ inference does not object to any information (including the state transition model) as long as the information is beneficial; the key is how to properly use information and knowledge of unknown quality.*

Overall, we cannot overuse any useful information, nor shall we omit any. The use of the information shall not only be based on its uncertainty (e.g. variance of the noise) but also on its credibility namely the matching rate to the truth. This is the starting point of the $O_2$ inference approach that is aimed to avoid suspicious/unnecessary assumptions, so it turns to seek more information from trustable sensors. It is a conservative albeit reliable solution. However, we are not mentioning issues other than the state estimation such as system identification or parameter estimation (see e.g. [19, 23]) where a filter might still be very useful and necessary.

### III.B  Nonlinear inversing bias

As stated, the inversing will often introduce biases (i.e. the expectation of the estimate is not equal to the true state) if the observation function $h_t(\cdot)$ is nonlinear, where the bias is state-dependent and highly depends on both the noise and the nonlinearity. Simply, a nonlinear conversion of a Gaussian distribution is no more Gaussian and therefore the situation can be very complicated. Generally, the larger the noise and the nonlinearity, the larger the bias/error. This has been recognized when converting polar/spherical measurements to Cartesian coordinates for the use of filters, see e.g. [38, 39]. To a degree, the converting bias/error can be approximately removed in an explicit/analytical form for simple noises (such as Gaussian noises). Significantly different to existing work, the $O_2$ inference does not assume the observation noises and therefore it works for the case of unknown and even time-varying observation noises. Hence, we (have to) omit this issue when the noise $v_t$ is unknown by setting it to be zero as shown in Eq. (7). If the observation noise is known, we propose to use a Monte Carlo simulation method to remove the inversing bias/error as follows. This is different to the explicit methods given in [38, 39] and the references therein and is only concentrated with the estimate-mean.

The idea is simply sampling a set of (random or even deterministic) samples from the noise distribution, $\{v_t^{(i)}\}, i = 1,2,\ldots I$ and use them separately in the inversing calculation of (6) as

$$\hat{x}_t^{(i)} = h_t^{-1}\left(y_t, v_t^{(i)}\right), s.t. v_t^{(i)} \sim p(v_t) \tag{26}$$

and we have the estimate given as the mean of these sample estimates as

$$\hat{x}_t = \frac{1}{I}\sum_{i=1}^{I} \hat{x}_t^{(i)} \tag{27}$$

For multi-dimensional models where dimensions are correlated (explicit), a huge number of samples might be needed to statistically remove the estimate bias.

Obviously, this Monte Carlo simulation is unbiased and will remove the bias caused by the nonlinear inversing, regardless the type of noises and the observation function. This again follows the Remark 6 that any useful information shall be used and can be beneficial; here, the information of the observation noise is used.

## III.C  Joint application of multiple unbiased sensors

Multi-sensor data fusion provides several primary advantages over data from a single sensor [24]. First, combining the observations of identical sensors (e.g., identical radars tracking a moving object) will result in improved estimate accuracy, assuming the data are combined in an optimal manner (as addressed in Case 1). Second, using the relative placement or motion of multiple sensors can improve the observability (to solve the so-called under-determined observation problem, see section III.D). But, multiple sensor-data fusion also face additional challenges such as sensor data correlation, inconsistency, etc. for which many studies can already be found e.g. [25] and which will not be discussed here. The joint application of multiple sensors via network (e.g. wireless sensor network) promises an $O_2$ estimate that theoretically converges to the true state as the number of unbiased sensors increase, i.e. the more unbiased sensors are used, the smaller the variance of their fused estimate is. Therefore, we have the following remark

**Remark 7** *The multi-sensor $O_2$ inference can achieve any level of estimation accuracy given an adequate number of independent unbiased sensors.*

Furthermore, as the number of sensors used increase, the multi-sensor data fusion will be able to distinguish the real observation of objects statistically from clutter affording the $O_2$ inference the same ability as a filter (but model-free). We will demonstrate this for the first time in the simulation provided in Section IV.C where the $O_2$ method is shown to be able to deal with multiple object filtering in a clutter environment.

The core ideas of filters and the multi-sensor $O_2$ inference are illustrated in Fig. 13 from an information fusion perspective. In this case, Advanced KF includes suboptimal KFs such as EKF/UKF/Cubature KF (see [2, 14] and references therein) and adaptive/robust KFs, while Advanced PF include suboptimal PFs and some enhanced PFs e.g. [4, 6, 15, 18]. The multi-sensor $O_2$ inference fuses multi-sensor data while the filter fuses sensor data with the prediction. This reveals their core difference as addressed: *whether shall the information transferred from history be trusted/used*?

The sensor fusion is different to the prediction-correction fusion in the sense that the former merely relies on unbiased sensors (as believed) that gives guaranteed information while the latter applies unguaranteed state transition assumption. In order to achieve the estimation accuracy required, the filter propagates the information of history observations to fuse with the newest observation of the state which can be viewed as a sensor-saving albeit computationally expensive solution. The filters can execute information fusion in the case of a single sensor while $O_2$ inference cannot. In contrast, the multi-sensor $O_2$ inference is arguably *computationally cheap but sensor-expensive*, which can achieve a desirable performance by using more sensors. The advantage/disadvantage of filters and multi-sensor $O_2$ inference is obvious and can be summarized as follows:

**Remark 8** *The filter aims to use as much as possible information for (sub) optimality (albeit computation expensive) but suffer from modeling/approximation errors; the multi-sensor $O_2$ is reliable (albeit sensor expensive), which does not employ any unguaranteed information and is computational cheap but may leave out some useful information.*

There is a trade-off. On the one hand, we want to maximally fuse all information to obtain optimality for estimation. On the other hand, we need to be very careful with the quality of information. Also, we need to consider the computation speed desired and the number of sensors available. In practice, one may be more interested in investigating sensors rather than filters, for better computing speed and reliability. We call this the "*rich man principle*"! *Give me more sensors, and I shall not need any filter[2].*

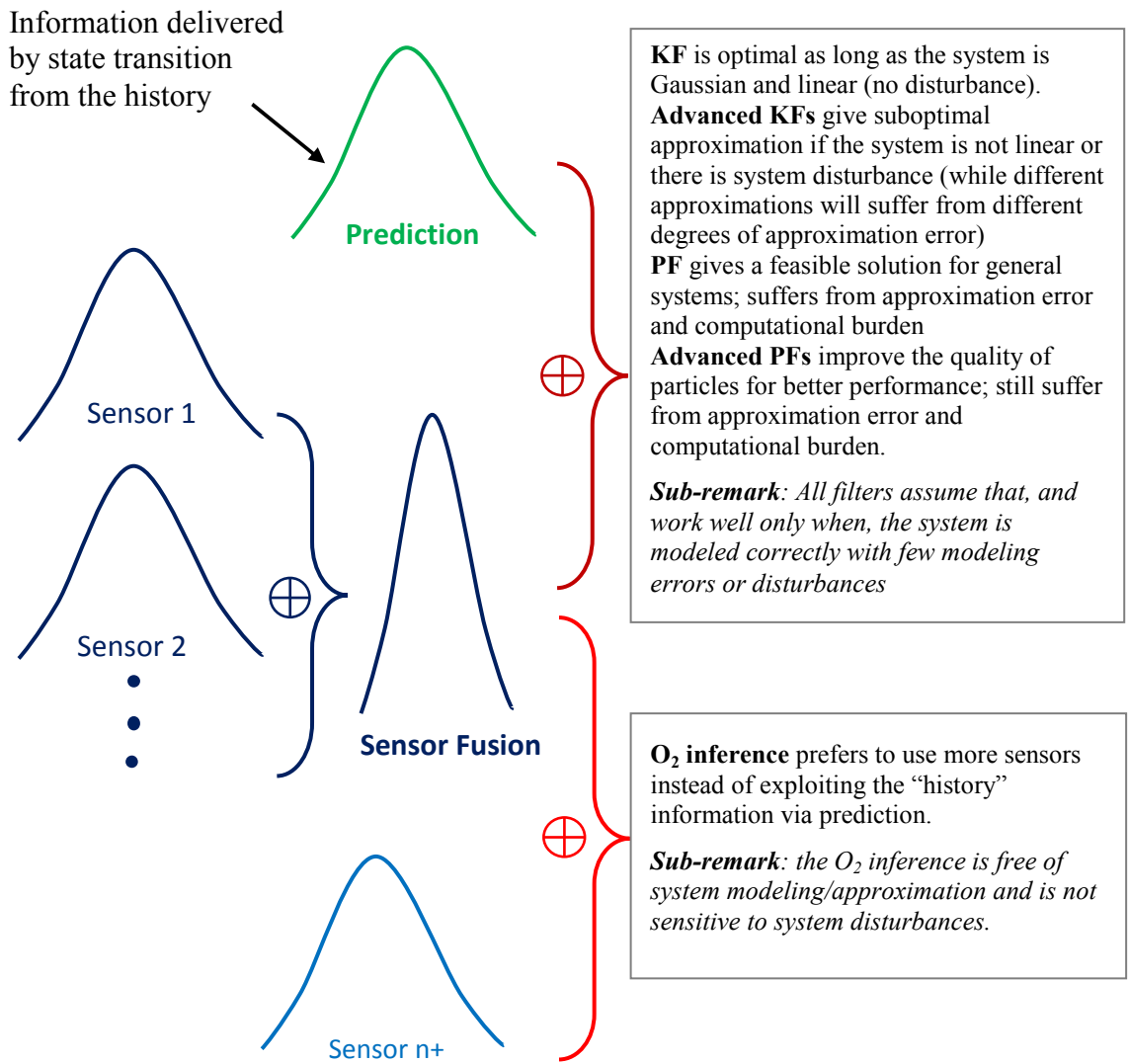

Fig.13 Information fusion involved in different filters and the multi-sensor $O_2I$ method
⊕ represents an information fusion operator

---

[2] *Aristotle: Give me a fulcrum, and I shall move the world.*

It is necessary to note, more and/or very accurate sensors might be not good for the filters. For example, a very small observation noise corresponds to a sharp likelihood function which can easily cause serious weight degeneration, or even failure, in the particle filter (all particles are of likelihood close to zero because the prediction does not match the observation; this will be shown by simulation in section IV.A and C). It does not make sense that more or more-accurate sensors would lead to a worse result.

**Remark 9** *More observations mean more information, which shall always be good for an estimator otherwise the estimator is problematic.*

### III.D  Irreversibility, over/under-determination and incomplete estimation

*1) Irreversibility*

The primary challenge for the application of the $O_2$ inference is from the irreversibility of the observation function, for which the direct inversing calculation is not applicable. Generally for an irreversible observation function, there exist multiple potential estimates $\hat{x}_{1,t}, \ldots \hat{x}_{m,t}$ corresponding to the same observation $y_t$. A favorable case is that the state is limited in a small state space based on a prior knowledge so that false estimates can be removed (the best is that only one potential estimate matches as the observation function is locally monotonous in that state space; see Section IV.B/C). More generally, the state dynamics knowledge (if known; or assumed as is done in the filter) can be explored. For this, one estimate that is closest to the prediction based on the previous estimate to $f_t(\hat{x}_{t-1})$ can be selected as the final estimate, i.e.

$$\hat{x}_t = \mathrm{argmin}_{x \in \{\hat{x}_{1,t}, \ldots \hat{x}_{m,t}\}} |x - f_t(\hat{x}_{t-1})| \tag{28}$$

In this solution, not only observation but also the state dynamics is used, namely the $O_2+$ inference. The sign estimate given by $\mathrm{sgn}(\hat{x}_t) = \mathrm{sgn}(f_t(\hat{x}_{t-1}))$ share the same idea.

As an alternative, we can explore multiple observations (by using multiple sensors in practice) on the same state, i.e. for the state $x_t$, we seek the best (approximate) solution for a set of observation equations as follows, given $n \geq m$

$$\begin{cases} y_{1,t} = h_{1,t}(x_t, v_{1,t}) \\ y_{2,t} = h_{2,t}(x_t, v_{2,t}) \\ \ldots \\ y_{n,t} = h_{n,t}(x_t, v_{n,t}) \end{cases} \tag{29}$$

where $y_{i,t}$ denotes the $i$ th observation, $v_{i,t}$ denote the $i$ th noise affecting the $i$ th observation equation $h_{i,t}$ corresponding to the $i$th sensor.

Then, the multi-sensor $O_2$ inference works by solving the equations (29) about the state $x_t$. To guarantee the equations is over/exact-determined, it is required the rank of the equations is larger or equal to $m$, e.g. it is better the sensors are located at different positions to avoid the singular problem.

Specifically, the $O_2$ inference Eq. (29) may be over-constrained/determined or under-constrained/determined, with regard to the dimensions of the state that are observed and the dimensions of the efficient observations. Generally, over-determination occurs when the total dimensions of all observations are more than the total freedoms of the state that are observed (given that the equations is non-singular), whereas under-determination occurs when the total dimensions of all observations are smaller than the total freedoms of the state that are observed. The under-determination belongs to one irreversible case in the view that observations are not enough to infer to the state.

  *2) Over-determination*

For an over-determined system, the $O_2$ inference shall use each independent subgroup of minimum observations to infer the estimates and finally fuse all estimates obtained according to their corresponding variance (i.e. KF-fusion in the Gaussian case) to get the final optimal estimate, where all observations shall be used equally. It is better to design an over-determination observation system so that the rank of equations (29) can be divided by $m$ without remainder (this is convenient for fusion). According to the KF-fusion, $N$ estimates of the same variance $P$ fused in the KF-manner will be equivalent to an estimate of variance $P/N$. The statistical advantage of multiple observations on the same scenario (i.e. over-determination) can also be used to distinguish the real observations from targets to clutter, i.e. clutter-filtering ability as stated.

For example, in the use of a laser radar for robot localization, as many as 180 scanning distances received at each scan may be available for estimating the 2-dimensional planner position of the robot. Ideally, two distance-data in a unique area of a map can infer one estimate of the position; one can therefore get as many as 90 estimates of the true state by using the $O_2$ inference on 90 pairs of distance-data. These 90 estimates can be fused according to their variances in the optimal manner.

  *3) Under-determination*

It is challenging to determine the state of an under-determined system, for which there often exist multiple potential estimates for the same observation. This is also challenging for filters. The solution that is workable in practice is to get more information, by either adding more sensors (of different observation functions) to get more observations in order to make the system exactly determined or even over-determined or by further exploring information from the state dynamics or others to remove suspicious estimates, as addressed with the irreversibility. For example, at least two bearing sensors are required to determine the planner positions of the state of targets. Here, the $O_2$ inference is carried out by solving the set of observation functions as shown in Eq. (29).

More efficient treatments are desirable for specified observation functions. It shall be avoided to design an under-determined observation system. Fortunately, thanks to the rapid development of sensors (lower price and higher quality), the observation system is

T. Li *et al*. Do we always need a filter? arXiv: 1408.4636

more possible to be over-determined in realistic applications. However, *for conservative reasons, we don't argue that the $O_2$ inference is applicable for all cases*.

4) *Estimation of the unobserved dimensions of the state*

**Remark 10** *The $O_2/O_2+$ approach is only able to directly estimate the dimensions of the state that have been observed, while the unobserved dimensions of the state shall be further inferred through the observed dimensions if they are related*.

The first-hand inference given by the observation inference might be an incomplete estimation of the state (depending on the definition of the state!). For the dimensions that are unobserved but desired, further inference based on their relationships with the observed dimensions is required, e.g. in the target tracking context, range and bearing observation are all defined on the position while the Doppler observation is defined on the velocity information. If only the position of an object is observed, the $O_2$ inference can only directly provide the position estimate but not any information about its velocity; the same occurs when only the velocity is observed. Here again, as highlighted, the state transition knowledge will be useful. The differentiation of the position is the velocity, and the differentiation of the velocity is the acceleration. Furthermore, the classification of a target may be determined based on the feature of its trajectory.

Note that it is the same story in filters where the unobserved dimensions of the state are also inferred by association with the observed dimensions. But, it seems impossible to estimate the dimensions of the state that are fully independent of the dimensions observed, no matter filters or the $O_2$ inference. Regarding the challenges faced by realistic problems, the $O_2$ inference may still be inapplicable.

## III.E  Fisher efficiency and Cramér-Rao bound

It is clear that under the condition that the observation is unbiased, the $O_2$ inference is an unbiased estimator. In the following we will study the efficiency of the $O_2$ inference based on Fisher information. For simplicity, we concentrate on the 1D (scale) observation function with an additive zero-mean Gaussian observation noise, for which the $O_2$ inference will output an estimate $\hat{x}_t \sim \mathcal{N}(x_t, \sigma^2)$ where $x_t$ is the real state (the mean of the estimate), $\sigma^2$ is the estimate variance depending on the sensor. The Fisher information provides a tool of measuring the amount of information that the estimate $\hat{x}_t$ carries about the unknown parameter $x_t, \sigma^2$, which is calculated based on the probability density (also known as the likelihood function) as follows, in the Gaussian case

$$f(\hat{x}_t; x_t, \sigma^2) = \frac{1}{\sigma\sqrt{2\pi}} e^{-\frac{(\hat{x}_t - x_t)^2}{2\sigma^2}} \tag{30}$$

For this normal distribution, the Fisher information matrix (here for parameter $x_t, \sigma^2$) contained in the random observation-based estimate $\hat{x}_t$ is

28T. Li *et al*. Do we always need a filter? arXiv: 1408.4636

more possible to be over-determined in realistic applications. However, *for conservative reasons, we don't argue that the $O_2$ inference is applicable for all cases*.

4) *Estimation of the unobserved dimensions of the state*

**Remark 10** *The $O_2/O_2+$ approach is only able to directly estimate the dimensions of the state that have been observed, while the unobserved dimensions of the state shall be further inferred through the observed dimensions if they are related*.

The first-hand inference given by the observation inference might be an incomplete estimation of the state (depending on the definition of the state!). For the dimensions that are unobserved but desired, further inference based on their relationships with the observed dimensions is required, e.g. in the target tracking context, range and bearing observation are all defined on the position while the Doppler observation is defined on the velocity information. If only the position of an object is observed, the $O_2$ inference can only directly provide the position estimate but not any information about its velocity; the same occurs when only the velocity is observed. Here again, as highlighted, the state transition knowledge will be useful. The differentiation of the position is the velocity, and the differentiation of the velocity is the acceleration. Furthermore, the classification of a target may be determined based on the feature of its trajectory.

Note that it is the same story in filters where the unobserved dimensions of the state are also inferred by association with the observed dimensions. But, it seems impossible to estimate the dimensions of the state that are fully independent of the dimensions observed, no matter filters or the $O_2$ inference. Regarding the challenges faced by realistic problems, the $O_2$ inference may still be inapplicable.

## III.E  Fisher efficiency and Cramér-Rao bound

It is clear that under the condition that the observation is unbiased, the $O_2$ inference is an unbiased estimator. In the following we will study the efficiency of the $O_2$ inference based on Fisher information. For simplicity, we concentrate on the 1D (scale) observation function with an additive zero-mean Gaussian observation noise, for which the $O_2$ inference will output an estimate $\hat{x}_t \sim \mathcal{N}(x_t, \sigma^2)$ where $x_t$ is the real state (the mean of the estimate), $\sigma^2$ is the estimate variance depending on the sensor. The Fisher information provides a tool of measuring the amount of information that the estimate $\hat{x}_t$ carries about the unknown parameter $x_t, \sigma^2$, which is calculated based on the probability density (also known as the likelihood function) as follows, in the Gaussian case

$$f(\hat{x}_t; x_t, \sigma^2) = \frac{1}{\sigma\sqrt{2\pi}} e^{-\frac{(\hat{x}_t - x_t)^2}{2\sigma^2}} \tag{30}$$

For this normal distribution, the Fisher information matrix (here for parameter $x_t, \sigma^2$) contained in the random observation-based estimate $\hat{x}_t$ is

28T. Li *et al*. Do we always need a filter? arXiv: 1408.4636

more possible to be over-determined in realistic applications. However, *for conservative reasons, we don't argue that the $O_2$ inference is applicable for all cases*.

4) *Estimation of the unobserved dimensions of the state*

**Remark 10** *The $O_2/O_2+$ approach is only able to directly estimate the dimensions of the state that have been observed, while the unobserved dimensions of the state shall be further inferred through the observed dimensions if they are related*.

The first-hand inference given by the observation inference might be an incomplete estimation of the state (depending on the definition of the state!). For the dimensions that are unobserved but desired, further inference based on their relationships with the observed dimensions is required, e.g. in the target tracking context, range and bearing observation are all defined on the position while the Doppler observation is defined on the velocity information. If only the position of an object is observed, the $O_2$ inference can only directly provide the position estimate but not any information about its velocity; the same occurs when only the velocity is observed. Here again, as highlighted, the state transition knowledge will be useful. The differentiation of the position is the velocity, and the differentiation of the velocity is the acceleration. Furthermore, the classification of a target may be determined based on the feature of its trajectory.

Note that it is the same story in filters where the unobserved dimensions of the state are also inferred by association with the observed dimensions. But, it seems impossible to estimate the dimensions of the state that are fully independent of the dimensions observed, no matter filters or the $O_2$ inference. Regarding the challenges faced by realistic problems, the $O_2$ inference may still be inapplicable.

## III.E  Fisher efficiency and Cramér-Rao bound

It is clear that under the condition that the observation is unbiased, the $O_2$ inference is an unbiased estimator. In the following we will study the efficiency of the $O_2$ inference based on Fisher information. For simplicity, we concentrate on the 1D (scale) observation function with an additive zero-mean Gaussian observation noise, for which the $O_2$ inference will output an estimate $\hat{x}_t \sim \mathcal{N}(x_t, \sigma^2)$ where $x_t$ is the real state (the mean of the estimate), $\sigma^2$ is the estimate variance depending on the sensor. The Fisher information provides a tool of measuring the amount of information that the estimate $\hat{x}_t$ carries about the unknown parameter $x_t, \sigma^2$, which is calculated based on the probability density (also known as the likelihood function) as follows, in the Gaussian case

$$f(\hat{x}_t; x_t, \sigma^2) = \frac{1}{\sigma\sqrt{2\pi}} e^{-\frac{(\hat{x}_t - x_t)^2}{2\sigma^2}} \tag{30}$$

For this normal distribution, the Fisher information matrix (here for parameter $x_t, \sigma^2$) contained in the random observation-based estimate $\hat{x}_t$ is

T. Li *et al*. Do we always need a filter? arXiv: 1408.4636

$$I = \begin{bmatrix} \frac{1}{\sigma^2} & 0 \\ 0 & \frac{1}{2\sigma^4} \end{bmatrix} \quad (31)$$

The Cramér-Rao Bound (CRB) given by the inverse of the Fisher information matrix provides a lower bound on estimation performance of any unbiased estimator [31] for a vector of non-random parameters. That is, the CRB for our case is

$$B_{O_2}(\hat{x}_t) = \begin{bmatrix} \sigma^2 & 0 \\ 0 & 2\sigma^4 \end{bmatrix} \quad (32)$$

The variance of the $O_2$ inference is $\sigma^2$ which is equal to the CRB on the state (the up-left element of matrix (32)). Thus, the $O_2$ inference is an *efficient* estimator for the state.

Obviously, the probability density function $f(\hat{x}_t; \theta)$ dominates the calculation of the Fisher information and the CRB, where $\theta$ is the parameter to estimate. For the $O_2$ inference, what known is only information from observations (nothing is assumed/used on the HMM and the prior background) and therefore estimates at different time are independent with each other. Correspondingly, the Fisher information matrix does not include the history/prior information. From this viewpoint, our approach pursues directly maximum likelihood estimation rather than maximum a posterior (MAP) estimation.

In contrast, the Bayesian CRB (BCRB) or posterior CRB [32] that is defined as the inverse of the Bayesian information matrix provides a lower bound for Bayes filters. It is based on the filtering posterior distribution and the calculation is generally of no closed-form expression for nonlinear systems. As such, a variety of alternative Bayesian bounds have been proposed, see e.g. [32]. We are not intended to detail them here. However, it is necessary to note that, most bounds are only applicable for *unbiased* estimator. Therefore the BCRB provided for filters only hold under the prerequisite that the filter is unbiased. Nevertheless, as addressed it is only under very properly system modeling (without any mismatching, error and biased approximation) and correctly known parameters that the filtering estimate is possibly guaranteed to be unbiased.

Last but not least, the $O_2$ method has extremely fast computing speed, which is highly preferable for real-life applications. This is because it is of the lowest computational complexity of all estimators e.g. in the manner of Bayesian information criterion[3] (detailed analysis is omitted here). Faster computing speed corresponds to smaller time-intervals between successive estimates, corresponding further to smaller process noise and lesser system disturbance [11]. This will not only obtain a more accurate estimate of the state in each scan[4] (due to less system disturbance and process noise between successive scans), but is also able to avoid/reduce observation redundancy[5] and thereby obtain more estimates. The availability of more, and more accurate, estimates in the same time-period will further provide a more accurate and smoother estimate of the continuous trajectory of the state. As such, computing speed is a critical factor to evaluate the performance of the estimator for in real-life applications. Further work is desirable to study the positive promotion of faster computing speed to estimation accuracy, for both the $O_2$ inference and filters.

---

[3] See e.g. http://en.wikipedia.org/wiki/Bayesian_information_criterion
[4] In the context of real time video tracking, faster computing speed corresponds to shorter processing time requirement for each frame, resulting in more frames the video steam can be divided in real time, lesser image difference between successive frames and lesser process noise [11]. Lesser process noise and object movement are positive for better tracking accuracy of a filter.

## IV    Simulations: $O_2$ vs filters

The biggest challenge for the filter is modeling error (including disturbance), which will almost surely prevent the fusion benefit as addressed. However, in our following simulations, we will employ exactly correct models and system noises for all filters to allow them to achieve the best possible performance. This is the most favorable situation for filters (otherwise if any mismodeling occur to the state transition function or the noises, the performances of the filters will be highly reduced).

Another limitation of the simulations is that, all estimators run in the same iterative frequency and receive the same amount of observations (as has been done in existing simulation). This however is unfair for the fast algorithm that does not need to "wait" for the slow one in realistic individual applications [11]. As addressed already, faster estimator will provide more, and more accurate, results in the same time period in real life implementations. In all the simulations, the $O_2$ inference is the fastest and has to *wait* for filters. This put the filters again in the favorable situation for comparison.

### IV.A    Single-observation $O_2$ inference

We will compare the $O_2$ inference with filters on another classic 1-demensional model that is also widely used since [5]. The state transition equation and the observation equation are given respectively as follows

$$x_t = \frac{x_{t-1}}{2} + \frac{25 x_{t-1}}{(1+x_{t-1}^2)} + 8\cos(1.2(t-1)) + u_t \tag{33}$$

$$y_t = \frac{x_t^2}{20} + v_t \tag{34}$$

where the process noise $u_t$ is zero-mean Gaussian $u_t \sim \mathcal{N}(0,Q)$ and the observation noise $v_t$ is also zero-mean Gaussian $v_t \sim \mathcal{N}(0,R)$. $Q=10, R=1$ are the default parameter setting in many publications including [5].

Inversing Eq. (34) after abandoning the noise term, the (biased) $O_2$ inference gives

$$\hat{x}_t = \pm\sqrt{20 \times y_t} \tag{35}$$

---

[5] In realistic use, the operating frequency of sensors can be faster than the iteration of the filter and thus, observations received are more than the handling ability of the filter, resulting in redundancy. In this case, faster computing speed indicates more utilization of observations and thus more estimates.

T. Li *et al*. Do we always need a filter? arXiv: 1408.4636Here, we explore three different ways to determine the sign of the estimate given by Eq. (35). The first uses one step of state transition function (default solution), the second uses the PF filtering result, and the third uses the true sign (although it is in fact unknown; here we assume there is one method that could well capture the sign of the true state or we are only interested in the absolute value of the estimate). They correspond respectively to the following three calculations

$$\hat{x}_t = \text{sgn}\left(\frac{\hat{x}_{t-1}}{2} + \frac{25\hat{x}_{t-1}}{(1+\hat{x}_{t-1}^2)} + 8\cos(1.2(t-1))\right)\sqrt{20 \times y_t} \quad (36a)$$

$$\hat{x}_t = \text{sgn}(\hat{x}_{t,PF})\sqrt{20 \times y_t} \quad (36b)$$

$$\hat{x}_t = \text{sgn}(x_t)\sqrt{20 \times y_t} \quad (36c)$$

Specifically, we will also apply the debiasing strategy given in Eq. (27) on (36c), i.e.

$$\hat{x}_t = \text{sgn}(x_t) \times \frac{1}{I}\sum_{i=1}^{I}\left(\sqrt{20 \times \left(y_t - v_t^{(i)}\right)}\right) \quad (36d)$$

where $I$ is the number of noise samples for debiasing and we set $I = 100$.

For comparison, the EKF, UKF (the unscented parameters are set as $\alpha = 1, \beta = 0, \kappa = 2$ as did in Section I, which however are by no means to be the best choice), auxiliary PF (APF) [40], Gaussian PF (GPF) [41] as well as the basic SIR PF have been implemented. The root mean square error (RMSE) is used and is defined as usual as follows

$$\text{RMSE}_1 = \sqrt{\frac{1}{M}\sum_{i=1}^{M}(x_{t,i} - \hat{x}_{t,i})^2} \quad (37)$$

where $M$ is the number of MC runs, $x_{t,i}$ is the true state at time $t$ of run $i$ and $\hat{x}_{t,i}$ is the state-estimate. To capture the average performance, 100 MC runs are executed with random re-initialization for each run. Each run consists of 100 time-steps.

When all PFs use 100 particles, the true state and estimates given by different filters are plotted in Fig.14, and the mean and variance of RMSE are given in Table II. Then, for a range of different number of particles from 20 to 500 used for the PFs, the mean RMSE and computing time of different filters and the $O_2$ inference are given in Fig. 15 and 16. Finally, for a range of different observation noise variances $R = [0.00001, 100]$, the mean RMSE of different filters (where all PFs use 100 particles) and the $O_2$ inference are given in Fig. 17. These results show:

1) The $O_2$ inference is extremely computing faster than all the filters except the $O_2$ inference with the use of SIR which needs the SIR to estimate the sign of the estimate.

2) Compared with others, the PFs (SIR, GPF and APF) do not make much difference with each other for this model. Specifically, a small observation variance is not always good for the PF no matter GPF, APF or SIR: when $R$ is reduced from 1 to 0.00001 or increase from 1 to 10000, the RMSE of the estimate increases. The best $R$ for them is around 1. When the observation noise variance is larger than 1, it is straightforward that the larger the noise is, the worse the filters are. But, a very accurate observation (e.g.

$R < 1$) corresponds to a sharp likelihood distribution, which can cause significant weight degeneracy and impoverishment (a particular problem of PFs), reducing the filter quality.

3) With the number of particles used increasing, the PFs will output better results (up to a stable level) but will also consume more time.

4) The unbiased $O_2$ inference performs the best of all the $O_2$ inference approaches.

Furthermore, we have the following findings:

1) With regard to the default $O_2$ inference, all the simulated filters outperform the $O_2$ inference except the EKF for $R < 10$ approximately while with regard to the $O_2$ inference with the PF for the sign of the estimate, all the filters outperform the $O_2$ inference for different observation noise except the EKF for $R < 1000$ and the UKF for $R \in [0.01, 10]$, approximately.

2) With regard to the $O_2$ inference with the correct sign (biased or unbiased), all the filters underperform the $O_2$ inference when the observation noise is small i.e. $R < 100$ approximately but is effective when $R > 100$ execpt the EKF, which seems always ineffective as long as $R < 10000$. This is because this model is highly nonlinear which is unsuitable for the application of the EKF.

The results show that, for this simulation model, the sign of the estimate affects the $O_2$ inference significantly as both Eq. (36a) and (36b) can estimate the sign wrongly due to the frequent switch of the sign between positive and negative as shown in Fig.14 (e.g. in some points, the estimates given by the naive $O_2$ inference are the wrong sign). In this aspect, this model is very challenging. The wrong choice of the sign of the estimate will significantly increase the $\text{RMSE}_1$. This is the reason why the default $O_2$ inference performs poorly. However, it is possible to find a method to estimate the sign of the state and, therefore, efforts should be made to do so, which might be more valuable than designing a filter for this model. We need to reiterate that the sign problem does not exist (at least not so significantly) in other models, such as target tracking where the state of interest is commonly bounded in a limited region (e.g. a known view field), see our simulation in section IV.C. Simply, if we define the RMSE[6] as follows

$$\text{RMSE}_2 = \sqrt{\frac{1}{M}\sum_{i=1}^{M}(|x_{t,i}| - |\hat{x}_{t,i}|)^2} \tag{38}$$

Then, the sign is no longer a problem. The default $O_2$ inference will perform the same as the $O_2$ Inference with correct sign.

Table II Performance of different estimators (100 MC runs)

|  | RMSE | |
|---|---|---|
|  | mean | variance |
| EKF | 15.193 | $9.47 \times 10^{-7}$ |
| UKF | 7.254 | 0.001 |
| SIR | 3.951 | 0.294 |
| GPF | 4.520 | 0.253 |
| APF | 4.207 | 0.448 |
| $O_2$ Inference | 16.243 | $1.4 \times 10^{-29}$ |
| $O_2$ Inference with PF for sign | 4.063 | 0.602 |
| $O_2$ Inference with correct sign | 1.391 | $5.5 \times 10^{-32}$ |
| Unbiased $O_2$ Inference with correct sign | 1.229 | $5.8 \times 10^{-4}$ |

---

[6] This metric is inspired by prof. Petar Djurić at Stony Brook University in our email conversations.

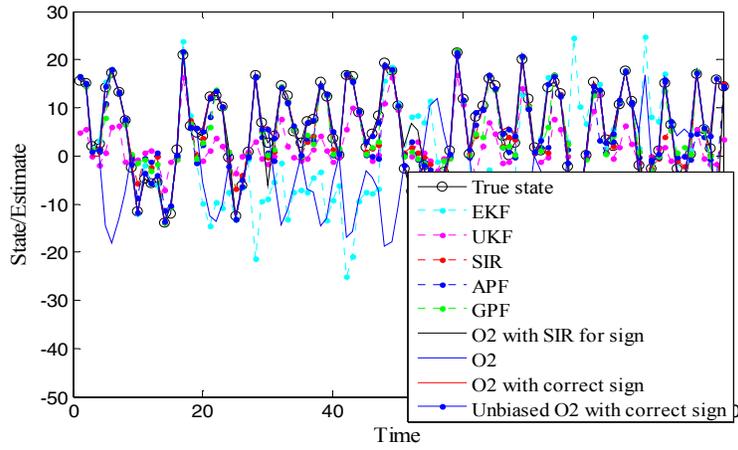

Fig.14  True state and estimates of different filters and the $O_2$ Inference

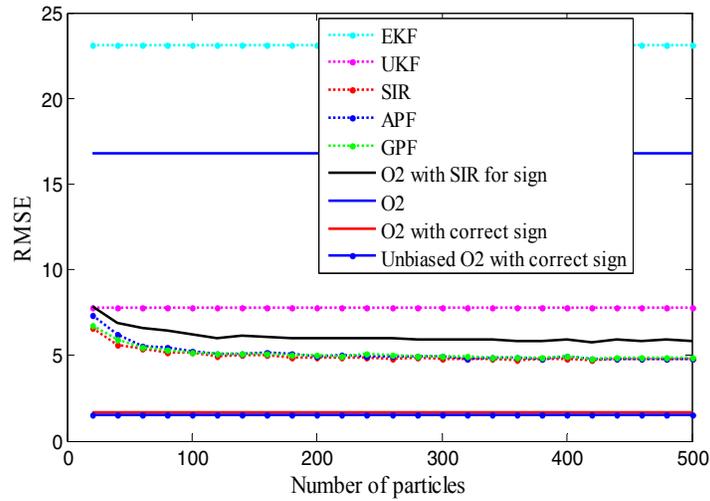

Fig.15  RMSE of different filters and the $O_2$ inference against the number of particles used in PFs

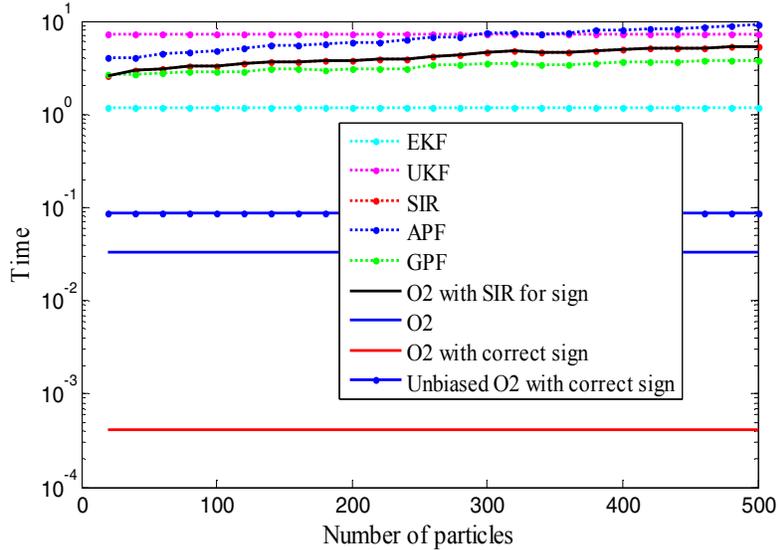

Fig.16  processing time against the number of particles used in PFs

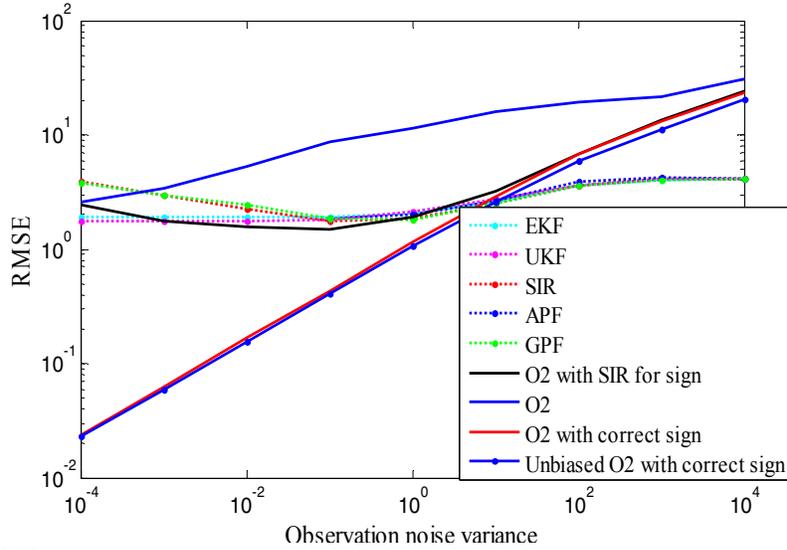

Fig.17  RMSE of different filters and the $O_2$ Inference against different observation noise variance

Furthermore, we revisit the simulation model given in Section I. For a range of different observation noise variances $R = [0.00001, 100]$ (other parameters are the same as given in Section I), the average RMSE (mean) of all the filters and the $O_2$ inferences is given in Fig.18. It can be seen that (*approximately*): when $R < 0.04$, all the filters are inferior to the $O_2$ inference; when $0.04 < R \le 1$, PF outperforms the $O_2$ inference while the other do not; when $2 < R \le 40$, PF, UKF and EKF outperform the (biased and unbiased) $O_2$ inference while the EKPF and UKPF do not (as addressed, emphasizing the observation in filter works only when the observation is pretty good); when $40 < R$, all filters used outperform the $O_2$ inference and simply become useful. This demonstrates again that filters can easily underperform the $O_2$ inference in certain cases.

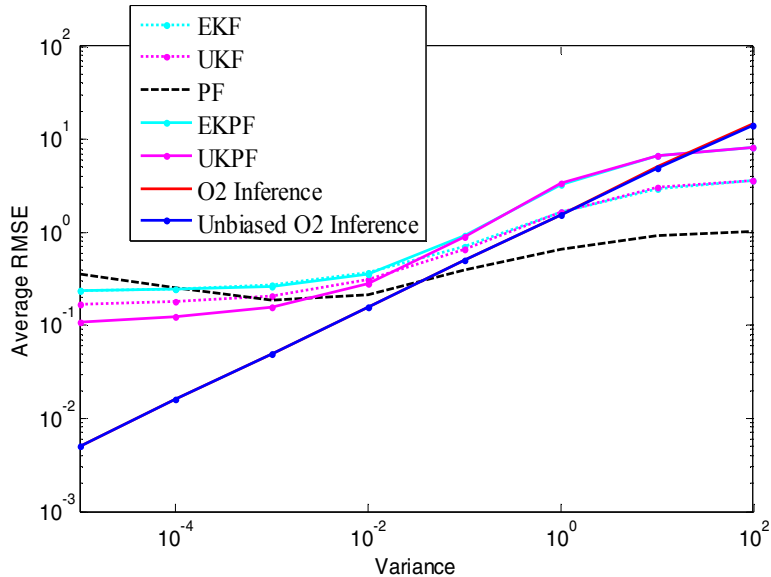

Fig.18 Average RMSE of different estimators of 60 steps×100 MC runs for different observation noise variances (R) on the simulation model (3~4) of Section I

## IV.B  Single sensor multi-target tracking

Multi-target tracking involves the joint estimation of the number of multiple objects and their states in the presence of clutter and noise given the observations from sensor(s). We will study the case of using one single sensor or multiple sensors separately.

In this simulation, we will show that the $O_2$ inference is workable as an estimate calculation method for an existing multi-target tracker namely the SMC-PHD filter (the sequential Monte Carlo implementation of the probability hypothesis density filter, also referred to the PF-SMC filter). Since the general multi-target tracking scenario contains miss-detection, target random birth and death and clutter, data-association/clutter-filtering algorithms are generally needed. Based on the random finite set and the point process theory, the PHD filter is established to propagate the first-order moment associated with the multi-target posterior through prediction and correction steps [17], which provides a clutter-filtering function while additional multi-estimate extraction from the joint-density of multiple targets is still required. In our case, three methods are employed for multi-estimate extraction. The first is the most common used *k*-means clustering [26]. The second is the state-of-the-art multi-expected a posterior (MEAP) estimator (see [18]). The third is the $O_2$ inference. That is, we apply three different estimate extraction methods based on the same output of a single SMC-PHD filter.

The true trajectories of targets are plotted in Fig.19 where the color distinguishes different birth models and each trajectory starts at '∆' and ends at '□'. As shown, new targets appear from four different areas following a Poisson RFS with intensity $\gamma_t = \sum_{i=1}^{4} r_{t,i} N(\cdot; B_i, Q)$, where $B_1 = [-1500,0,250,0,0]$, $B_2 = [-250,0,1000,0,0]$, $B_3 = [250,0,750,0,0]$, $B_4 = [1000,0,1500,0,0]$, $Q$ =diag $([50,50,50,50,6 \times \pi/180]^T)^2$ and $r_{t,1} = 0.02, r_{t,2} = 0.02, r_{t,3} = 0.03, r_{t,4} = 0.03$. The survival probability of a target is $p_S = 0.99$. The target state variable $x_t = [\tilde{x}_t, \omega_t]^T$ consists of the planar position and velocity $\tilde{x}_t = [p_{x,t}, \dot{p}_{x,t}, p_{y,t}, \dot{p}_{y,t}]$ and the turn rate $\omega_t$. The nearly constant turn-rate state transition model can be written as

$$\tilde{x}_t = F(w_{t-1})\tilde{x}_{t-1} + Gw_t, w_t = w_t + \Delta u_{t-1} \tag{39}$$

where

$$F(\omega) = \begin{bmatrix} 1 & \frac{\sin \omega \Delta}{\omega} & 0 & -\frac{1-\cos \omega \Delta}{\omega} \\ 0 & \cos \omega \Delta & 0 & -\sin \omega \Delta \\ 0 & \frac{1-\cos \omega \Delta}{\omega} & 1 & \frac{\sin \omega \Delta}{\omega} \\ 0 & \sin \omega \Delta & 0 & \cos \omega \Delta \end{bmatrix}, G = \begin{bmatrix} \frac{\Delta^2}{2} & 0 \\ \Delta & 0 \\ 0 & \frac{\Delta^2}{2} \\ 0 & \Delta \end{bmatrix}$$

$w_{t-1} \sim N(\cdot; 0, \sigma_w^2 I), u_{t-1} \sim N(\cdot; 0, \sigma_u^2 I), \Delta = 1\text{s}, \sigma_w = 15\text{m/s}^2$ and $\sigma_u = \pi/180\text{rad/s}$.

The range-bearing observation region is the half disc of radius 2000m. The detection probability of a target is $p_{D,t}(x) = 0.95\mathcal{N}([p_{x,t}, p_{y,t}]^T; 0, 6000^2 I_2)/\mathcal{N}(0; 0, 6000^2 I_2)$. If detected, the observation is a noisy range and bearing vector given by

$$z_t = \begin{bmatrix} r_t \\ \theta_t \end{bmatrix} = \begin{bmatrix} \sqrt{p_{x,t}^2 + p_{y,t}^2} \\ \arctan\left(\frac{p_{x,t}}{p_{y,t}}\right) \end{bmatrix} + v_t \tag{40}$$

where $v_t \sim N(\cdot; 0, R_t)$, with $R_t = \text{diag}([\sigma_r^2, \sigma_\theta^2]^T)$, $\sigma_r = 5$m, $\sigma_\theta = \pi/180$rad/s.

The optimal sub-pattern assignment (OSPA) metric [27] is used to evaluate the estimation accuracy. A big OSPA distance indicates low estimation accuracy. For arbitrary finite subsets $X = \{x_1, x_2, \ldots, x_m\}$ and $Y = \{y_1, y_2, \ldots, y_n\}$ where $m, n \in \mathbb{N}_0 = \{0,1,2,\ldots\}$, the OSPA metric of order $p$ between $X$ and $Y$ is defined as (if $m \leq n$)

$$\bar{d}_p^{(c)}(X, Y) = \left(\frac{1}{n}\left(\min_{q \in \Pi_n} \sum_{i=1}^m d^{(c)}(x_i, y_{q(i)})^p + c^p(n-m)\right)\right)^{1/p} \tag{41}$$

where $d^{(c)}(x, y) = \min(c, d(x, y))$, the cut off value $c > 0$ and $d(x, y)$ is the Euler distance. $\bar{d}_p^{(c)}(X, Y) = \bar{d}_p^{(c)}(Y, X)$ if $m \geq n$ and $\bar{d}_p^{(c)}(X, Y) = 0$ if $m = n = 0$. The parameters used for the OSPA are $c = 100, p = 2$.

For the filter implementation, 1000 particles per expected target are used and the total number of particles is hard-limited to be not less than 600. Clutter is uniformly distributed over the region with an average rate of $r$ points per scan, i.e. $\kappa_t = r/2000/\pi$. We apply a large range of clutter rate and run 100 Monte Carlo trials. The standard $k$-means algorithm (runs up to 50 iterations if the algorithm does not converge), MEAP and the O$_2$ method are independently applied on the same PDF for multi-estimate extraction.

Instead of clustering all of the particles whose distribution is a fusion of the predicted PHD and the observation according to the PHD updater, the MEAP method extracts the estimates based on the posterior likelihood distribution according to individual observations. The main role of the PHD filter is to identify the observations that are more possibly coming from real target and to propagate the particles. Then, the MEAP estimate is calculated based on these identified observations, as follows.

$$x_{z_t}^{EAP} = \frac{\sum_{i \in \Xi(z_t)} g_t(z_t|x_t^{(i)}) w_{t|t-1}^{(i)} x_t^{(i)}}{\sum_{i \in \Xi(z_t)} g_t(z_t|x_t^{(i)}) w_{t|t-1}^{(i)}} \tag{42}$$

where $z_t$ is the observation at time $t$ that is identified from the real target according to the PHD filter, $g_t(z_t|x_t^{(i)})$ is its likelihood of the particle $x_t^{(i)}$, $w_{t|t-1}^{(i)}$ is the predicted weight of particle from previous iteration, $\Xi(a)$ is the finite set of particles that are associated to observation $z_t$ based on the near and nearest neighbor (NNN) principle; and for simplicity, one can just use all particles $\Xi(z_t) = [1, L_t]$ where $L_t$ is the total number of

T. Li *et al*. Do we always need a filter? arXiv: 1408.4636particles for updating at time $t$ (they do not make much difference when targets are well distant). The idea of MEAP [18] is to formulate the multi-estimate extraction problem approximately as a set of parallel single-observation single-estimate extraction problems, which intrinsically emphasizes the observation in its calculation. More straightforwardly, one can simply calculate the estimate from the observation without knowing the noise, i.e. inversing Eq. (40) after taking off the unknown noise $v_t$, we have ($[r_t, \theta_t]^T = z_t$, where $z_t$ is the identified observation as stated)

$$\begin{bmatrix} p_{x,t} \\ p_{y,t} \end{bmatrix} = \begin{bmatrix} tan\,(\theta_t)\sqrt{\frac{r_t^2}{1+\theta_t^2}} \\ \sqrt{\frac{r_t^2}{1+\theta_t^2}} \end{bmatrix} \qquad (43)$$

We reiterate that this nonlinear conversion contains a bias because of the nonlinear inversing of the noise, although generally it is insignificant when the target is far from the sensor. As shown in [39] and the reference therein, it is not easy to explicitly remove the bias fully. But, the proposed Monte Carlo method given in (27) is theoretically able and computationally easy to remove the estimate-mean bias. However, we assume that the observation noise is not known in the $O_2$ inference and we will not apply any debiasing strategy. This accommodates the most general case that no noise information is feasible and allows a maximal degree of modelling-free for the approach.

Since it is known that the tracking scenario is in the area of $p_{y,t} > 0$, the sign of the state in *y*-dimension is always positive, while in *x*-dimension it is the same with $tan\,(\theta_t)$. Therefore, this inverse function does not have a sign problem. As remarked, the $O_2$ inference is only able to estimate (directly) the dimensions of the state that are observed (here, it is the position of targets) but not the dimensions that are not observed (i.e. velocity and turn rate). To note, the velocity and turn rate are not independent to the position but have a differentiation and integration relationship with the position. They can be further inferred from the differentiation of the position (with the condition that the turn rate is nearly constant) if necessary. The same occurs in the SMC-PHD filter (but in a soft way) where the position of the state is also obtained by integration of the velocity with regard to the turn rate and the previous state.

It can be found that both the MEAP and the $O_2$ method are highly based on individual observations for estimate calculation. The MEAP additionally explores the prediction information as it uses the weight $w_{t|t-1}^{(i)}$ which is propagated according to the state transition (39) while the $O_2$ method is purely based on observation, which does not care about what the state transition function is. In the following, we will compare these two methods with the standard *k*-means clustering method that aims to partition all the particles into *k* clusters in which each particle belongs to the cluster with the nearest mean, which is, however, computationally difficult (NP-hard).

First, we set the clutter rate $r = 10$. The observations of range and bearing in one trial are given in Fig.20 where the true trajectories in the observation space are given in the blue line while the observations are plotted in the black circle. The mean OSPA and computing time given by different methods are separately given in Fig.21. The performance of different methods are given in Table III. The results show that the $O_2$ inference, which is computationally fastest, achieves better estimation accuracy than the standard *k*-mean clustering method but slightly lesser than the MEAP method.

For a range of clutter rate $r = 0 \sim 30$, the results of the $O_2$ and MEAP as compared to the standard clustering based SMC-PHD filter are given in Fig. 22, which demonstrates again the $O_2$ method outputs the fastest computational speed and more accurate results than the *k*-means method, although it is slightly worse than the MEAP method. The results show that the observation-oriented methods, namely MEAP and $O_2$ inference, have achieved better results than the *k*-means. However, as previously mentioned, the prediction information is not fully useless in this proper assumed models as it can be explored to assist the observation-based inference of the state, which is the reason why the MEAP method that uses the particles propagated according to the state transition model gets more accurate results than the $O_2$ method (although not use so much as the clustering method). In other words, the MEAP method sits between traditional clustering approaches and the $O_2$ method, in the sense of the degree of the observation/prediction used. This can be viewed as a new fusion between prediction and correction that is not strictly the Bayes manner. However, it is still unclear what the best fusion balance is between assumed state transition and observation for any dynamic state space model; more theoretical work is expected in this regard.

However, the next simulation will show that with a large number of sensors available, the multi-sensor data fusion can serve as a clutter-filtering algorithm that will enable the $O_2$ method work independently for multi-target estimation in the clutter environment. Also, the use of massive sensors will alleviate the nonlinear inversing bias significantly.

Table III  The performance of different MEE methods when *r*=10

|      | *k*-means | MEAP     | $O_2$ inference |
|------|-----------|----------|-----------------|
| OSPA | 48.101    | 33.865   | 35.077          |
| Time | 0.0075    | 0.000066 | 0.000028        |

T. Li *et al*. Do we always need a filter? arXiv: 1408.4636

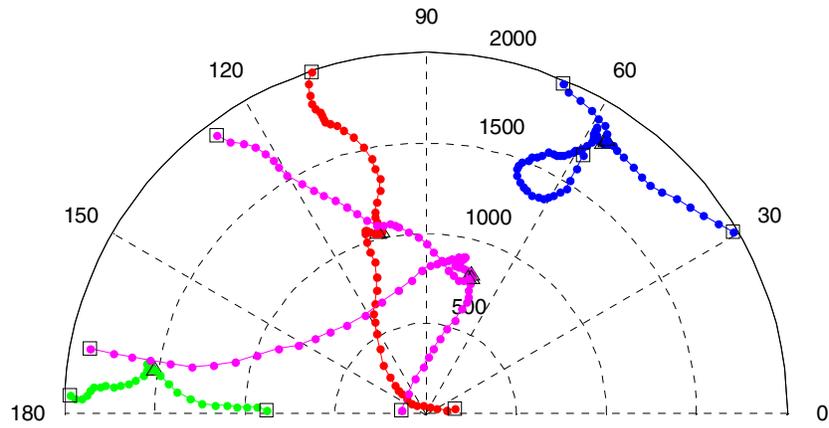

Fig. 19 Trajectories of targets born in four different areas

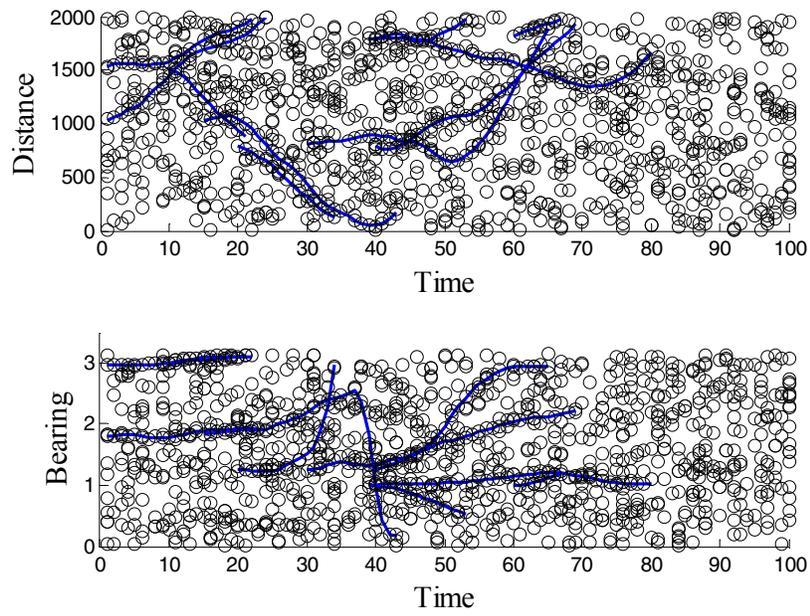

Fig. 20 Range-bearing observations (black o) and the true trajectories of targets (blue line) when $r = 10$

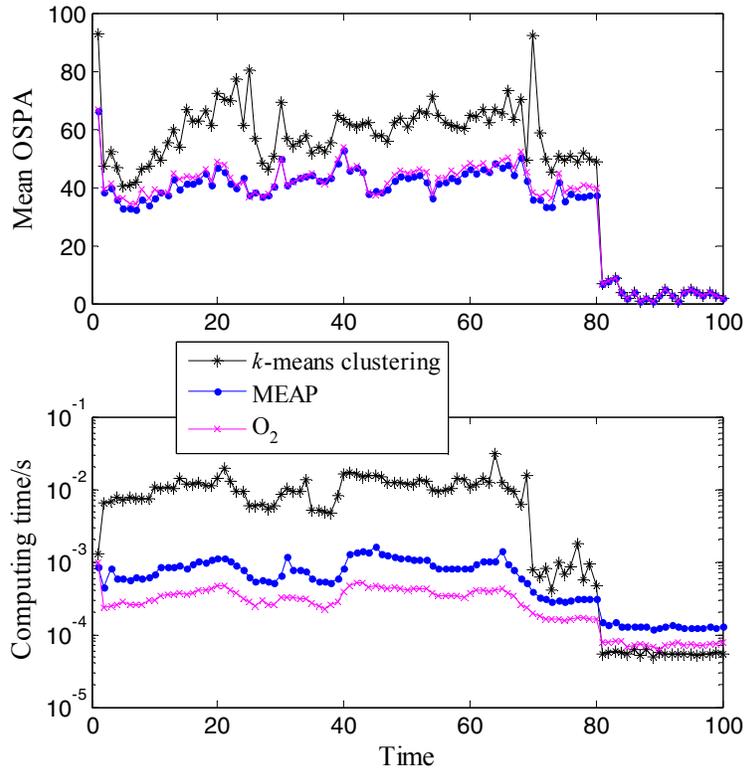

Fig. 21 Mean OSPA and computing time of different estimators when $r = 10$

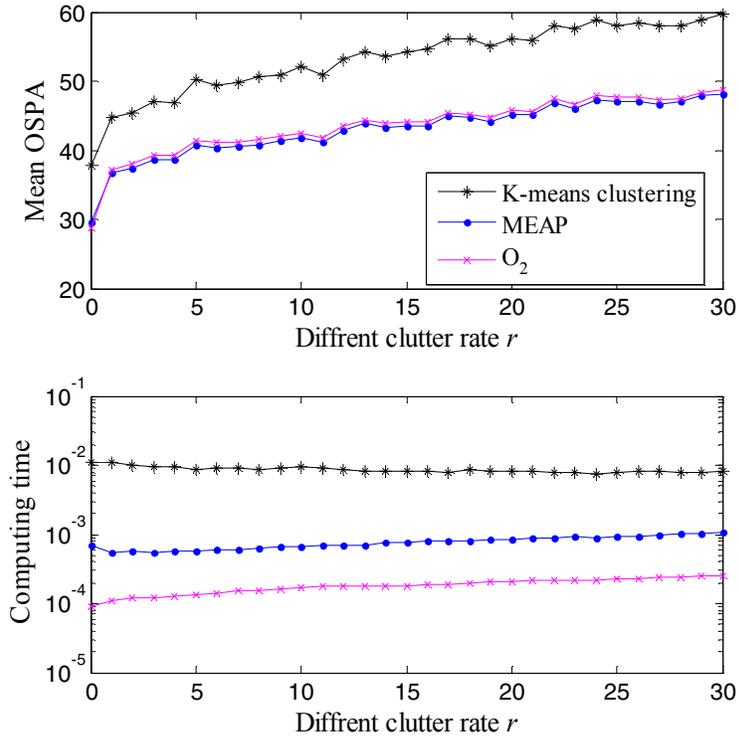

Fig. 22 Mean OSPA and computing time of different estimators for $r = 0 \sim 30$

T. Li *et al*. Do we always need a filter? arXiv: 1408.4636

## IV.C  Massive-sensor multi-target tracking

Based on the same scenario as described in the last simulation of Section IV.B (specifically we only apply $r=10$), we use multiple/massive independent active (range-bearing) sensors that are of the same observation function (40) and are located at the same planner position for simplicity; this corresponds to the situation in which these sensors are arranged in the same planner position but at different altitudes. This is not mandatory as one can put these sensors at different positions and they can be of a different quality, but then their observation scope may be different. There is always a way to fuse these sensors. Compared with the previous simulation, lower quality sensors are used in this case that have observation noises $v_t \sim N(\cdot; 0, R_t)$, with $R_t = \text{diag}([\sigma_r^2, \sigma_\theta^2]^T)$, $\sigma_r = 20$m, $\sigma_\theta = \pi/90$rad/s. In this simulation, different number of sensors will be used.

We apply the O$_2$ method independently without any assistance from the SMC-PHD filter. In other words, *the O$_2$ method does not need to assume anything about the clutter and target model (e.g. appearing, moving and disappearing)*. Therefore, the O$_2$ method is insensitive to the uncertainty and change of the system model (and so don't need to care about maneuver problem). What is needed in the O$_2$ method is only the observation model. It is necessary to note, this is based on the very general assumption that the observations of one target are subject to a unimodel distribution while the clutter is scattered (e.g. randomly, uniformly) independently to targets. The principle can be explained as follows based on the visible simulation data.

Fig.23 gives the scanning observations of six sensors in one trial for $t = 40$. Mapping the observations of a total of ten sensors into the same state space (as shown in Fig.24), we will find that the observations (to be more precisely, estimates) will be of high density in the area containing targets, and of low density in other regions with no target. Based on this, we can confirm target existence in the area of high observation density as long as the observation noise is unbiased and the detection probability is relatively high. The observations of different sensors lying in the same high-density area are more likely from targets and as such, can be used together to extract estimates in the O$_2$ method. In order to do so, a proper clustering method that is sensitive to the data density, such as DBSCAN (Density-based spatial clustering of applications with noise[7]), will be helpful but here we propose a more straightforward threshold-based method as follows.

Denoting the total number of sensors as $N$ and the number of the sensors whose view fields cover the target $i$ as $N_i$, the detection probability of sensor $s$ on target $i$ can be denoted as $p_{D,s}(i) \leq 1$ which is usually a function on the distance between the sensor and the target. Denoting target $i$ reports roughly $n_i$ observations in all the sensors, we have

$$E(n_i) = \sum_s^{N_i} p_{D,s}(i) \leq N_i \tag{44}$$

---
[7] See e.g. http://en.wikipedia.org/wiki/DBSCAN

Since the target detection probability and the view field of the sensors is approximately deterministic for any given point-target, $n_i$ can be fully calculated based on the state of the targets. Therefore, we treat $n_i$ as a known parameter in this paper. The assumption (A.3) indicates that $n_i \approx (0.7 \sim 1) N_i$. This parameter is critical to identify the number of potential targets in the following clustering approach.

---

**Algorithm 1**    Clutter filtering based on clustering

---

1) Apply the $O_2$ inference on all the sensor data as addressed in Section 2, obtaining undistinguished data-points (including state-estimates of targets and false alarms) in the same space.

2) Calculate the distances between any two data-points from different sensors in the state space. Data-points from different sensors will be identified as connected if their distance is smaller than a threshold $d = l \times \sigma_v$ (where $\sigma_v$ refers to a rough estimate of the (largest, if different) standard deviation of the observation noise that is mapped in the state space, and we use a scaling parameter $l \in [1,4]$).

To note here, the parameter $d$ corresponds to the average distance between two data-points that are drawn from the same distribution under a confidence level $l$. Here if $\sigma_v$ is unknown, the threshold $d$ does not need to be very accurate and can actually be created through unsupervised learning about the data point set since the distances between data points from the same target will be significantly smaller than others. We will address this separately in our future work.

3) Since data-points in the same cluster can be from a single target or multiple close targets, a detection of the number of data-points in each cluster shall be applied to distinguish isolated targets from close-targets. Here, another threshold $p$ is needed to give the average number of data-points in a single cluster that contains target $i$ only. It shall be designed with respect to $N_i$ as defined in Eq. (44) e.g. $p_i = 0.8 \times N_i$ in our simulation. The static parameter 0.8 is scalable for fine adjustment.

3.1) If a data-point has been connected with more than $p_i$ but smaller than $2 \times p_i$ other data-points, the data-point and its connections will be identified from a single target, forming a sub-cluster to extract one estimate using Kalman fusion or simply calculating the mean.

3.2) If a data-point has been connected with more than $k \times p_i$ (but smaller than $(k+1) \times p_i$ where $k \geq 2$) other data-points, that data-point and its connections are identified from multiple close targets. Then, these closely connected data-points will be partitioned into $k+1$ groups based on their proximities in the state space, each of which has approximately $p_i$ but no more than $N_i$ data-points and will fuse them to extract one state-estimate.

To note here, the state-estimates of close-distributed targets, inferred from the same sensor shall be clustered into different groups even they are closely distributed in the space. Therefore, there shall be basically only one estimate in each sub-cluster that is from the same sensor. The obtained sub-clusters will have approximately equivalent number of data-points. In our current application, there are few targets ($k \leq 3$) moving closely, therefore making the partitioning of the cluster relatively easy. Close-target is also a very challenging problem for the traditional filter-based multi-target tracker [22].

---

We compare the multi-sensor O₂ method with the aforementioned MEAP SMC-PHD filter. There are several ways to incorporate different sensors to work for the SMC-PHD tracker that primarily differ from one another with regard to how the information fusion is implemented [16]. First, we apply a naive track-to-track (T2T) fusion that is to run the SMC-PHD filters separately for different sensors, and fuse their estimates eventually. In particular, the overestimation of the number of targets of one filter with regard to the average of others will be simply eliminated. Secondly, we apply a single "super" sensor based SMC-PHD filter where the single sensor has a much lower observation noise $v_t \sim N(\cdot; 0, R'_t)$, where $R'_t = \text{diag}([\sigma'^2_r, \sigma'^2_\theta]^T) = R_t/N$, i.e. $\sigma'_r = \frac{20}{\sqrt{10}}$m, $\sigma'_\theta = \frac{\pi}{90\sqrt{10}}$rad/s. This corresponds to an observation accuracy that is equivalent to the Kalman fusion of $N$ sensors used in the O₂ inference and the T2T SMC-PHD filter. This is referred to as observation fusion and tracking (OFT).

First, we set $N = 10$ for each of the three methods, i.e. ten sensors for the O₂ inference and the T2T SMC-PHD filter. The average results of the estimates of the number of targets and the mean OSPA of different filters over 100 Monte Carlo runs are given in Fig.25 and Fig.26. The average results over 100 steps×100 MC runs are summarized in Table IV where the time refers to one iteration of the filter. Although the multi-sensor O₂ method does not need any information about the target/clutter, it can still estimate the number of real targets with acceptable accuracy and has produced even better results (smaller OSPA) than the multi-sensor T2T SMC-PHD filter. This clearly demonstrates that the multi-sensor O₂ inference has a good clutter-filtering ability. To the best of our knowledge, this is the first time to filter clutter without a filter and without any prior knowledge of the target/clutter model.

We iterate that the correct assumption of the target and clutter model and relevant parameters required by the SMC-PHD filters that is the same to the ground truth is unavailable in real-life problems; then, the filters will not give as good a result, although the O₂ inference will be the same since it does not rely on these model at all. However, we will show that even in the case where the system is perfectly known by the filters, the multi-sensor O₂ method can still perform better than the filter with the number of sensors increasing. Using correctly assumed knowledge of the system, the OFT SMC-PHD filter performs better than the T2T SMC-PHD filter, indicating that the multi-sensor T2T solution implemented is not optimal. We will further compare the better OFT SMC-PHD filter (using MEAP estimator) with the O₂ inference under different $N$ (i.e. a different number of sensors used in the multi-sensor O₂ approach and a different corresponding observation noise used in the OFT SMC-PHD filter).

Fig. 27 gives the mean OSPA results obtained by the O₂ and the OFT SMC-PHD filter against different $N$. The results show that with an increase of $N$, the O₂ will surely get more reliable and accurate results, but this is not guaranteed with the SMC-PHD filter. The SMC-PHD filter gets the best accuracy when approximately $N = 10$. An observation noise that is too large (small $N$) is not good for all methods while a very small observation noise corresponds to a sharp likelihood distribution and thereby may cause significant sample impoverishment in PF (this might not appear to be significant in the Gaussian mixture implementation of the PHD filter [28]); both cases will cause the reduction of the performance of the particle filter. This simply exposes the advantage of the $O_2$ method, which guarantees reliable and consistent improvement in performance with the number of unbiased sensors used increasing.

Overall, the simulation has clearly demonstrated that 1) the multi-sensor $O_2$ inference has the filtering ability that is able to distinguish real targets from false alarms, as well as to estimate the number and state of targets in a clutter environment; 2) with an increasing number of sensors, the accuracy of the multi-sensor $O_2$ method will statistically increase.

Table IV The performance of different trackers when $r=10$, $N=10$

|      | MEAP SMC-PHD (OFT) | MEAP SMC-PHD (T2T) | $O_2$ inference (10 sensors) |
|------|--------------------|--------------------|------------------------------|
| OSPA | 31.5482            | 40.3143            | 36.9991                      |
| Time | 1.8905             | 19.771             | 1.1224                       |

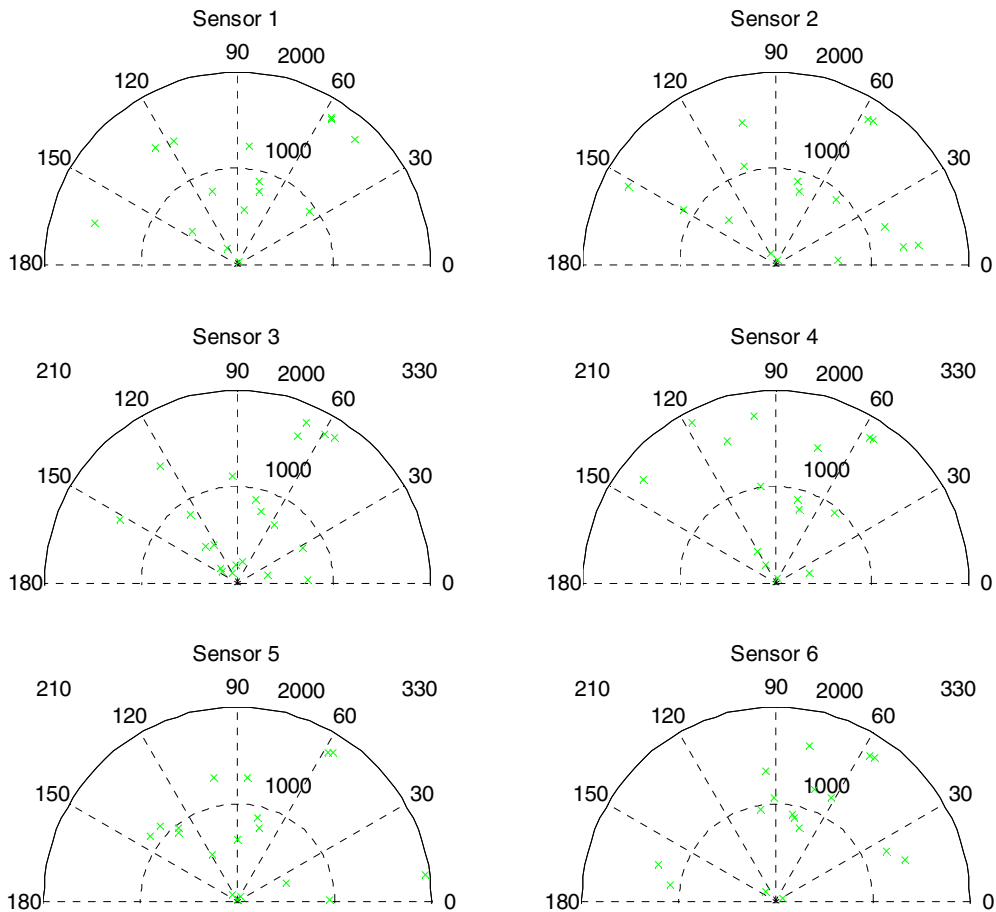

Fig. 23 Individual observations of six sensors at time $t = 40$ when $r = 10$

T. Li *et al*. Do we always need a filter? arXiv: 1408.4636

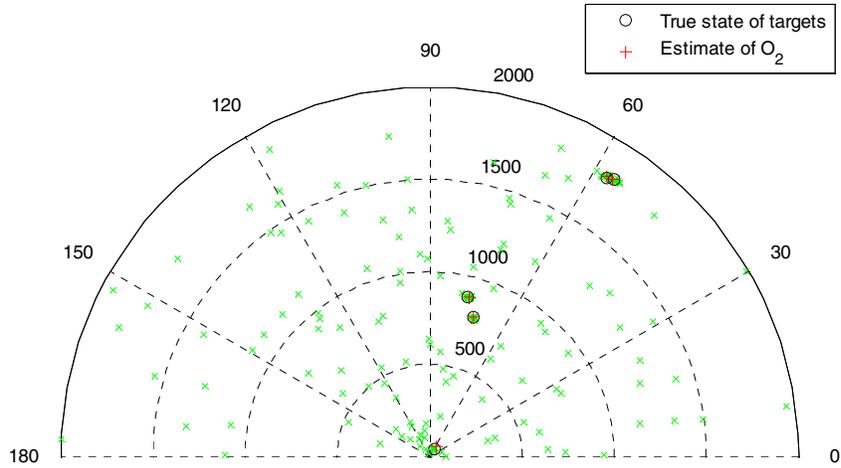

Fig.24 Observations of ten sensors (green x), true states (black o) and the $O_2$ estimates (red +) at time $t = 40$ when $r = 10$

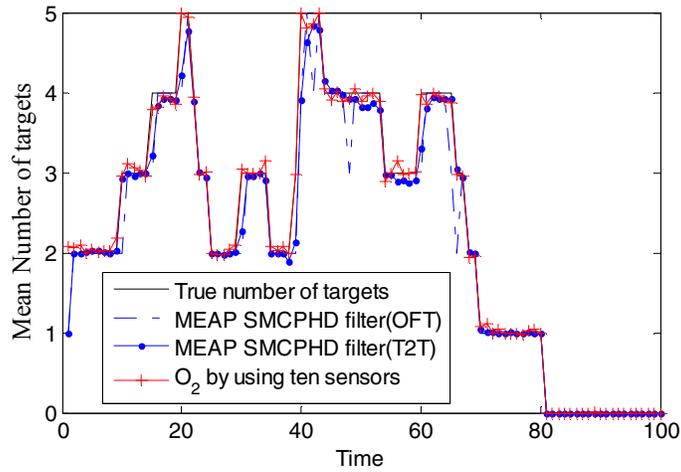

Fig. 25 Mean estimated number of targets given by different estimators over 100 Monte Carlo runs

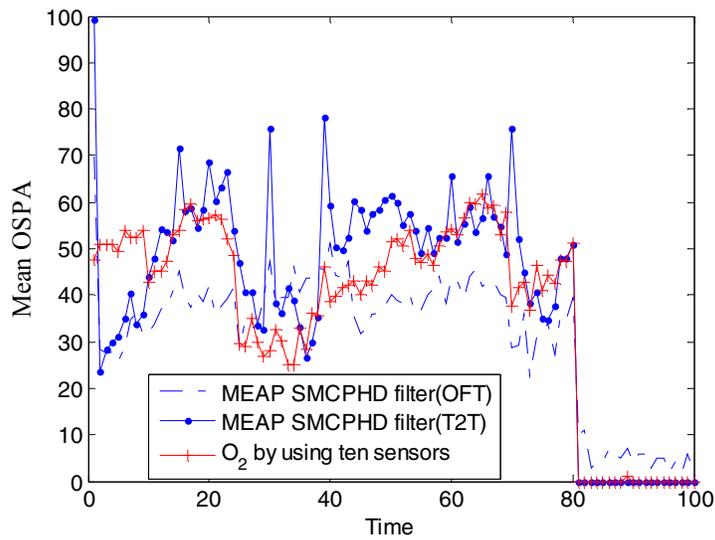

Fig. 26 Mean OSPA of different estimators over 100 Monte Carlo runs

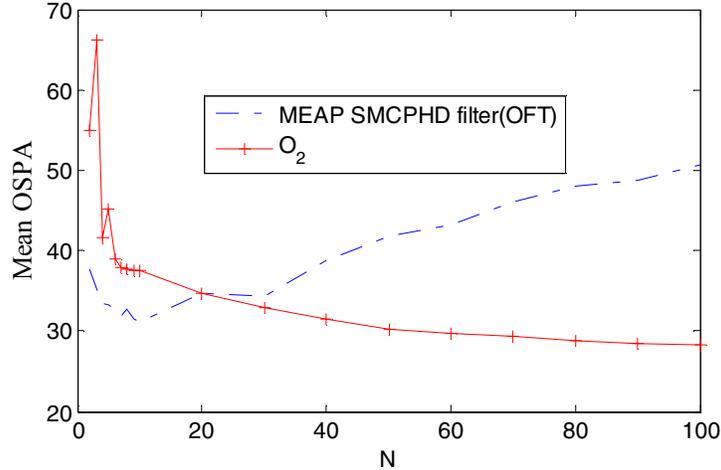

Fig. 27 Mean OSPA of 100 steps×100 MC runs of the $O_2$ inference and the OFT SMC-PHD filter for different $N$.

## IV.D  Ghost tracking

In this case we consider a particular scene in which the motion model of multiple "ghost" targets and the system noises are completely unknown and are possible time varying. The ghost targets can appear/disappear anywhere anytime in the scene jointly, adjacently or solitarily and they may split, merge or cross each other, just to name a few. Therefore, it is challenging and even impossible to use any existing filter-based tracker. In such a situation, traditional estimators are simply inapplicable unless equipped with strong maneuver handling ability (but will still be questionable) and will not be considered here. However, we can still apply multi-sensor $O_2$ inference as long as we receive observations of these targets' state and we know the observation function.

The trajectories of these ghosts over the view region $[-100, 100] \times [-100, 100]$ are given in Fig.28. These trajectories start at different times and exist for different lengths as shown in Fig.29. These trajectories are in fact randomly generated according to a constant velocity model, near constant velocity model, near constant turn-rate, high noisy constant turn models and fully static target model. However, the estimator does not know this. Clutter is uniformly distributed over different regions with an average rate of $r = 10$ points per scan (this is also not unknown in the $O_2$ estimator). Denoting the position of the ghost as $[p_{x,t}, p_{y,t}]^T$, the observation equation is given by

$$z_t = \begin{bmatrix} z_{x,t} \\ z_{y,t} \end{bmatrix} = \begin{bmatrix} p_{x,t} \\ p_{y,t} \end{bmatrix} + v_t \qquad (45)$$

with $v_t$ is an unknown unimodal noise. For simulation only, we use $v_t = [u_{x,t}, u_{y,t}]^T$, $u_{x,t}$ and $u_{y,t}$ as mutually independent zero-mean Gaussian noise with a large variance 25; this is not unknown in the $O_2$ estimator.

The observation equation (45) could correspond to the case of cameras reporting the pixel position of the target in the image (based on some image/video processing technologies). There is quite less information provided in such a MODE tracking scene. The given information is only about the observation function (45) and the observations. The unbiased $O_2$ inference has

$$\begin{bmatrix} p_{x,t} \\ p_{y,t} \end{bmatrix} = \begin{bmatrix} z_{x,t} \\ z_{y,t} \end{bmatrix} \tag{46}$$

To note, this inversing calculation is linear and is unbiased.

We use ten sensors for the multi-sensor $O_2$ inference and set the clutter rate $r = 10$. Correspondingly, all the observations from ten sensors, the true state positions and the estimates by $O_2$I are plotted in Fig.30. The average results of the estimates of the number of targets and the mean OSPA over 100 Monte Carlo runs are given in Fig.31 and Fig.32. The average OSPA over 100 steps×100 MC runs is 30.128, which is arguably a good result with regard to the average number of targets and clutter and the parameter $c = 100, p = 2$ used for OSPA. This demonstrates again that the multi-sensor $O_2$ inference is not only an independent MODE estimator but can also be very reliable and accurate, even in the highly unknown and time-varying clutter environment.

Finally, we point out that additional strategies are required for the $O_2$ inference for further tasks, such as track continuity [21, 26], which associates estimates generated at different time-instants that belong to the same target to form the complete trajectory/track of separate targets. For this, as remarked already, state transition information as well as data association algorithms will be helpful and necessary. For example, the estimates obtained in different time-instants can be smoothed according to known target dynamics, and then connected to form continuous trajectories.

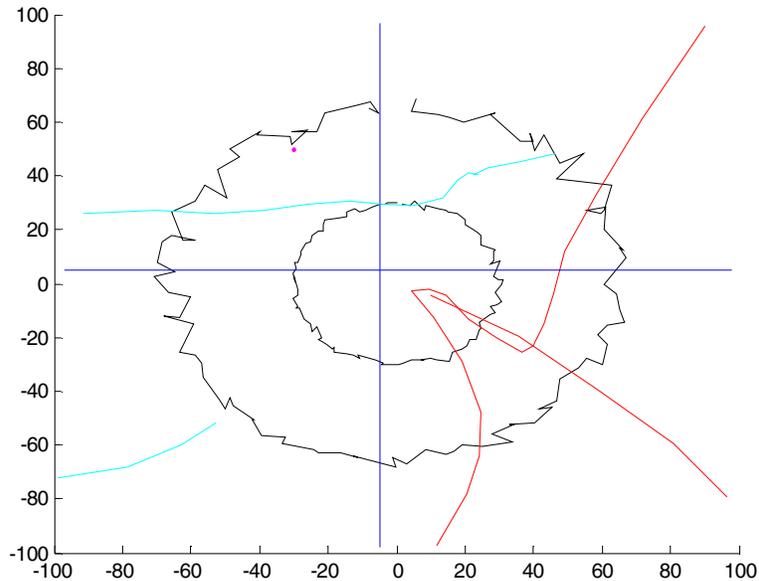

Fig.28 Trajectories of ghost targets

T. Li *et al*. Do we always need a filter? arXiv: 1408.4636

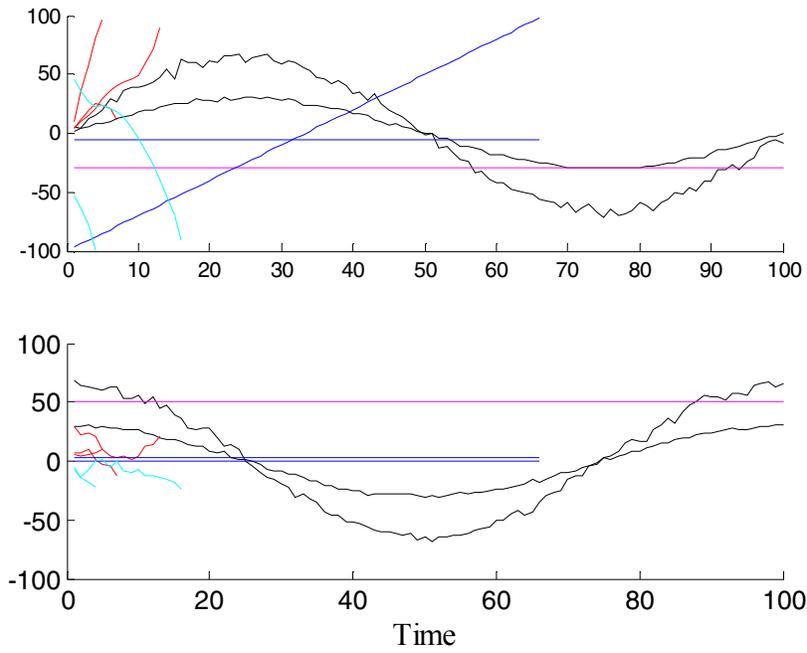

Fig. 29 True trajectories of targets in $x - y$ dimension separately (the color is consistent with Fig.28)

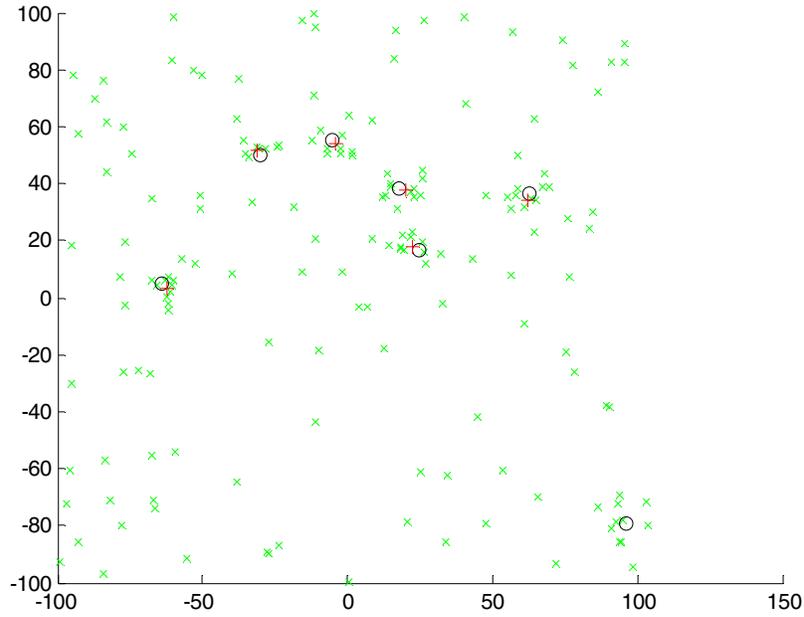

Fig. 30 Observations (green "x") of ten cameras, true targets ("o") and estimates (red "+") at time $t = 16$

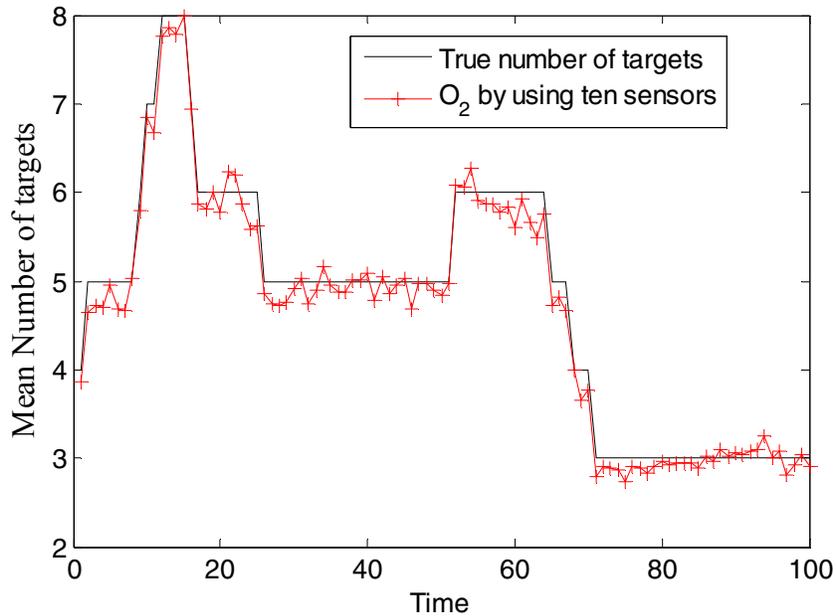

Fig. 31 Mean estimated number of targets over 100 Monte Carlo runs

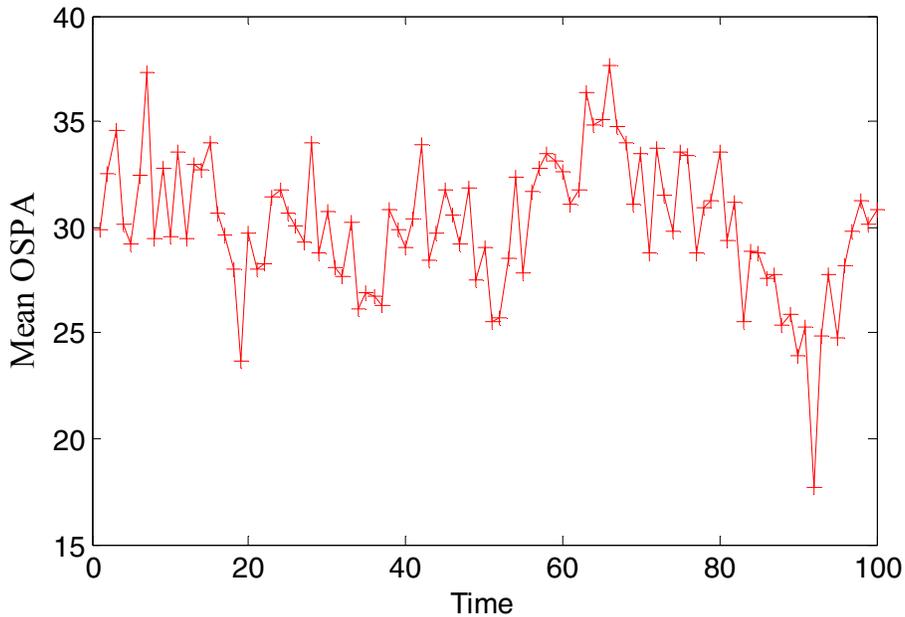

Fig. 32 Mean OSPA over 100 Monte Carlo runs

## V    Conclusion

The essence of discrete-time filters of the prediction-correction format is the fusion of the history information and the new observation to make (sub) optimal estimation. Good results require correct and accurate models and few system disturbances/approximate errors otherwise the filter does not guarantee a benefit as compared to the straightforward observation-only ($O_2$) inference and plus ($O_2$+). We have quantitatively investigated when and why the Kalman and particle filters do not give a more accurate estimation than the $O_2$ inference, especially for the general case that the observation is unbiased. A large number of simulations on several typical models have demonstrated the theory findings. Based on this, we emphasize that a discrete-time filter shall only be applied for state estimation when it gives at least a better estimate than the $O_2$ inference (if applicable). Of the extreme computing speed, the performance of the $O_2$ inference has identified a benchmark to assess the effectiveness of filters where a filter is effective only when it at minimum outperforms the $O_2$ inference statistically in accuracy. More promisingly, the $O_2$ inference is a hardware-oriented solution that benefits directly from the rapid development of sensors and their embedded data processing algorithms. These allow us to reassess the realistic state estimation problem, for which we might consider more from the hardware point of view.

The advantages of the multi-sensor $O_2$ inference for independent reliable multi-state estimation are noted. Although multi-sensor data fusion has been acknowledged and investigated intensively within the target-tracking context, it is for the first time explicitly exploited as an independent clutter-filter and employed with the $O_2$ inference for general multi-target tracking. Given enough number of sensors, the multi-sensor $O_2$ inference can theoretically meet any level of estimation accuracy required (given proper debiasing technology for nonlinear observation model) and can itself filter clutter by fusing the observations from different sensors. The $O_2$ inference can also work with data-association or clutter-filtering algorithms to deal with complicated cases. Specially, the multi-sensor $O_2$ inference can work independently for the very challenging environment where the knowledge about the target and clutter models is quite limited and the target motion is of high maneuver. Future work will include analytical performance assessment of the O2 inference for different types of models and noises, generating continuous trajectories of separate targets (used further for prediction of the state), online unsupervised learning for the parameters required in the clustering method for massive (distributed and/or heterogeneous) sensor-based MODE.

## Acknowledgments

The authors have consulted and received comments from several excellent researchers who work in the field of tracking, estimation and adaptive control including prof./Dr. Yu-Chi (Larry) Ho, Huimin Chen, Miodrag Bolić, Petar, Djurić, Mahendra Mallick, Lyudmila S. Mihaylova, Zhengling Yang, Lingji Chen, Genshe Chen, Xin Tian, Bin Jia, Quan Pan et al. The authors acknowledge their insights and discussion.

Tiancheng Li's work is supported by the Excellent Doctorate Foundation of Northwestern Polytechnical University, China and by the Postdoctoral Fellowship of the University of Salamanca, Spain.